\documentclass[usegraphicx,usenatbib]{mn2e}
\usepackage{color}
\input{colordvi.tex}

\date{\today}

\begin{document}
\twocolumn
\title[Second-Order NG from Minkowski Functionals of WMAP Data]{Limits on Second-Order Non-Gaussianity from Minkowski Functionals of WMAP Data}
\author[Hikage et al.]{Chiaki~Hikage$^1$\thanks{hikage@kmi.nagoya-u.ac.jp},
  Takahiko~Matsubara$^{1,2}$ \\
\\
$^1$  Kobayashi-Maskawa Institute, Nagoya University, Nagoya 464-8602, Japan \\
$^2$  Department of Physics, Nagoya University, Nagoya 464-8602, Japan}
\maketitle

\begin{abstract}
We analyze non-Gaussianity (NG) due to the primordial bispectrum and
trispectrum using cosmic microwave background temperature maps of WMAP
7-year data.  We first apply the perturbative formulae of Minkowski
functionals (MFs) up to second-order NG derived by
\citet{Matsubara:10}, which enable us to give limits on cubic NG
parametrized with $\tau_{\rm NL}$ and $g_{\rm NL}$ as well as various
types of quadratic NG parametrized with $f_{\rm NL}$. We find no
signature of primordial NG in WMAP 7-year data, but give constraints
on the local-type, equilateral-type, orthogonal-type $f_{\rm NL}$:
$f_{\rm NL}^{\rm (loc)}=20\pm 42$, $f_{\rm NL}^{\rm (eq)}=-121\pm
208$, and $f_{\rm NL}^{\rm (ort)}=-129\pm 171$, respectively, and
$\tau_{\rm NL}/10^4=-7.6\pm 8.7$, and $g_{\rm NL}/10^5=-1.9\pm
6.4$. We also find that these constraints are consistent with the
limits from skewness and kurtosis parameters which characterize the
perturbative corrections of MFs.
\end{abstract}

\section{Introduction}
\label{sec:intro}
Non-Gaussianity (NG) in primordial curvature perturbation is a key
observational probe to study physics in early Universe. Single
slowly-rolling scalar inflation model predicts too small NG to be
observed \citep{Salopek90,Falk93,Gangui94,Acq03,Mal03,Bartolo06}.
Variants of the simple inflationary models generate much higher levels
of NG: multiple fields \citep{LM1997,Lyth03}; modulated reheating
scenarios \citep{DGZ2004}; warm inflation
\citep{GuBeHea02,MossXiong:07}; ekpyrotic model
\citep{Koya07,CS2007,Buch07}.
These NGs have been constrained using the cubic-order statistics
(i.e., bispectrum) from CMB temperature maps of WMAP
\citep{Komatsuetal:11,Creminelli:07,YadavWandelt:08,Curtoetal:11}.

So far NG analysis has been mainly focused on the measurement of
primordial bispectrum. Primordial trispectrum, the next higher-order
term, also provides an important probe to differentiate inflation
models. Trispectrum is generally described into two different kinds of
connected parts and their amplitudes is commonly parametrized by
$g_{\rm NL}$ and $\tau_{\rm NL}$ \citep[e.g,][]{OkamotoHu:02}. 
So-called local-type NG is described as the nonlinear correction in Bardeen's
curvature as follows:
\begin{equation}
\Phi=\phi+f_{\rm NL}(\phi^2-\langle\phi^2\rangle)+g_{\rm  NL}\phi^3...,
\label{eq:local}
\end{equation}
\citep[e.g.,][]{KogoKomatsu:06} where $\phi$ is an auxiliary Gaussian
variable and $\Phi$ is related to the curvature perturbation $\zeta$
as $\Phi=3/5\zeta$. In this local-type model, $\tau_{\rm NL}$ is equal
to $36f_{\rm NL}^2/25$. \citet{SuyamaYamaguchi:08} find that all
classes of multi-inflation models satisfy the inequality condition
$\tau_{\rm NL}\ge 36f_{\rm NL}^2/25$ derived from the Cauchy-Schwarz
inequality. The relation between $\tau_{\rm NL}$ and $f_{\rm NL}$ is
therefore another powerful probe of inflation models
\citep[e.g.,][]{STYY:10}.  Current WMAP constraints on $\tau_{\rm NL}$
and $g_{\rm NL}$ are obtained by several groups: $-5.6 <g_{\rm
  NL}/10^5<6.4$ (95\%C.L.)  from the N-point probability density
function \citep{VielvaSanz:10}; $-12.4<g_{\rm NL}/10^5 <15.6$
(95\%C.L.) using trispectrum \citep{Fergussonetal:10}; $-0.6<\tau_{\rm
  NL}/10^4<3.3$ (95\%C.L.) and $-7.4<g_{\rm NL}/10^5<8.2$ (95\%C.L.)
using kurtosis power spectrum \citep{Smidtetal:10}.  Planck is
expected to reach $\Delta\tau_{\rm NL}\sim 560$
\citep{KogoKomatsu:06}.

Full analysis of trispectrum is computationally very expensive because
their number of configurations are enormous. Minkowski functionals
(MFs) are another powerful tool to constrain NGs from the aspect of
morphological properties of the density structure. MFs and
polyspectrum such as bispectrum and trispectrum are very different
statistics. MFs are real-space statistics, whereas the polyspectrum is
defined in harmonic space. MFs naturally incorporate the information of
all orders of polyspectra and thus they are complementary to the
commonly used measurement using polyspectra.  According to the
perturbative formulae of MFs derived by \citet{Matsubara:10}, the
higher-order NG originated from primordial trispectrum is
characterized with four kurtosis parameters, which is the summation
over trispectrum with different weights on tetrahedron
configuration. The computation of kurtosis parameters is much less
expensive compared to the trispectrum estimator. MFs are
model-independent statistics and hence they have a potential to
serendipitously find NGs due to unknown sources that has not been
explored. Consistency check using different estimators from the
standard one is thereby important to obtain more robust results.

In this paper, we first apply the perturbative formulae of MFs upto
2nd-order derived by \citet{Matsubara:10} to WMAP 7-year data of CMB
temperature anisotropies. We give new constraints on $\tau_{\rm NL}$
and $g_{\rm NL}$ as well as local, equilateral, and orthogonal types
of $f_{\rm NL}$s. This paper is the extension of the previous work to
give limits on the local-type $f_{\rm NL}$ from MFs of WMAP 3-year
data \citet{Hikageetal:08} and BOOMERanG data \citet{Natolietal:10}.

The paper is organized as follows: in section \ref{sec:theory}, we
review the perturbative formulae derived by \citet{Matsubara:10}. In
section \ref{sec:simulations}, we compare the perturbative predictions
with numerical simulations. We apply the perturbative formulae to the WMAP data and constrain
different NG parameters in section \ref{sec:observations}. Section
\ref{sec:summary} is devoted to the summary and discussions.

\section{Perturbative Formulae of Minkowski Functionals for CMB Temperature Maps with Primordial NG}
\label{sec:theory}

\subsection{Minkowski Functionals for Gaussian Fields}
Minkowski functionals (MFs) have been used to characterize the
morphology of a given density field
\citep{MeckeBuchertWagner:94,SchmalzingBuchert:97}. In 2-dimensional
field such as CMB temperature maps, three MFs are defined: area
fraction ($V_0$), circumference ($V_1$), and Euler characteristics
($V_2$). We measure them for the area whose temperature anisotropy
$f\equiv \Delta T/T$ normalized by the standard deviation
$\sigma_0\equiv\langle f^2\rangle^{1/2}$ is larger than a threshold
value $\nu$.  In Gaussian temperature maps, the $k$-th MF $V_k^{\rm
  (G)}$ is given by
\begin{equation}
V_k^{\rm (G)}(\nu) = A_k\exp\left(-\frac{\nu^2}{2}\right)H_{k-1}(\nu),
\end{equation}
where $H_k(\nu)$ represent the $k$-th Hermite polynomials.  The
amplitude $A_k$ is given by
\begin{equation}
\label{eq:mfamp}
A_k=\frac1{(2\pi)^{(k+1)/2}}\frac{\omega_2}{\omega_{2-k}\omega_k}
\left(\frac{\sigma_1}{\sqrt{2}\sigma_0}\right)^k,
\end{equation}
where $\omega_k\equiv \pi^{k/2}/{\Gamma(k/2+1)}$.
The standard deviation $\sigma_0$ and that of the first derivative
$\sigma_1\equiv \langle|\nabla f|^2\rangle^{1/2}$ are written as a sum
of the power spectrum $C_l$:
\begin{equation}
\label{eq:var}
\sigma_j^2\equiv \frac1{4\pi}\sum_l(2l+1)\left[l(l+1)\right]^j C_l W^2_l,
\end{equation}
where $W_l$ represents the smoothing kernel determined by the pixel
and beam window functions and any additional smoothing. Here we use
Gaussian kernel $W_l=\exp(-l(l+1)\theta^2/2)$ and $\theta$ denotes the
smoothing angular scale. Table \ref{tab:scale} lists HEALPix pixel
number $N_{\rm side}$ (the total pixel number is $12N_{\rm side}^2$)
and the maximum multipole number $l_{\rm max}$ at different values of
$\theta$. We choose the values of $N_{\rm side}$ and $l_{\rm max}$ so
that the effects of pixel window and high frequency cut be small.  As
discussed later, measuring MFs of CMB maps with different smoothing
scales is important to extract configuration dependence contained in
primordial bispectrum and trispectrum.

\subsection{Perturbative Formulae in Weakly Non-Gaussian Fields}
\citet{Matsubara:03} has applied multivariate Edgeworth expansion
theorem to derive perturbative formulae of MFs for a general
field. \citet{Matsubara:10} extended his analysis and derived
2nd-order correction of MFs on CMB temperature maps. According to
their works, the MFs of a weakly NG field (i.e. $\sigma_0\ll 1$) are
written upto the 2nd-order term of $\sigma_0$ as
\begin{eqnarray}
V_k(\nu)&=&V_k^{\rm (G)}(\nu)+A_ke^{-\nu^2/2}\Delta v_k(\nu), \\
\Delta v_k(\nu) &=& v_k^{(1)}(\nu)\sigma_0+v_k^{(2)}(\nu)\sigma_0^2.
\end{eqnarray}
In the following subsection, we review the perturbative formulae for a
CMB temperature map.
\subsubsection{First-order perturbation}
The 1st-order perturbation terms of MFs are characterized by three
skewness parameters:
\begin{eqnarray}
\label{eq:delmf_pb}
v_k^{(1)}(\nu) =
\frac{S}{6}H_{k+2}(\nu)-\frac{S_{\rm I}}{2}H_k(\nu)-\frac{S_{\rm II}}{2}H_{k-2}(\nu),
\end{eqnarray}
with
\begin{equation}
\label{eq:skewness}
S\equiv\frac{\langle f^3\rangle}{\sigma_0^4},~~~
S_{\rm I}\equiv\frac{f^2\langle \nabla^2 f\rangle}{\sigma_0^2\sigma_1^2},~~~
S_{\rm II}\equiv\frac{2\langle |\nabla f|^2 \nabla^2 f\rangle}{\sigma_1^4}.
\end{equation}
The skewness parameters are written as the sum of the bispectrum with
different weights of triangle configurations:
\begin{equation}
S_A=\frac{1}{4\pi\sigma_0^4}\sum_{l_1,l_2,l_3}I_{l_1l_2l_3}\tilde{S}_{Al_1l_2l_3}B_{l_1l_2l_3}
W_{l_1}W_{l_2}W_{l_3},
\end{equation}
where
\begin{eqnarray}
\tilde{S}_{l_1l_2l_3}&=&1, \\
\tilde{S}_{{\rm I}~ l_1l_2l_3}&=&-\frac{\{l_1\}+\{l_2\}+\{l_3\}}{6q^2}, \\
\tilde{S}_{{\rm II}~l_1l_2l_3}&=&\frac{1}{12q^4}[\{l_1\}^2+\{l_2\}^2+\{l_3\}^2 \nonumber \\
& & -2(\{l_1\}\{l_2\}+\{l_2\}\{l_3\}+\{l_3\}\{l_1\})],
\end{eqnarray}
with $q=\sigma_1/\sqrt{2}\sigma_0$ and $\{l\}\equiv l(l+1)$, and 
\begin{equation}
I_{l_1l_2l_3}\equiv\sqrt{\frac{(2l_1+1)(2l_2+1)(2l_3+1)}{4\pi}}
\left(
\begin{tabular}{ccc}
$l_1$ & $l_2$ & $l_3$ \\
0 & 0 & 0
\end{tabular}
\right).
\end{equation}
Note that the proportional factor of the skewness parameters are
different from the parameters $S^{(i)}$ used in
\citet{Matsubara:03,Hikageetal:06} and their relations are $S=S^{(0)},
S_{\rm I}=-4S^{(1)}/3$ and $S_{\rm II}=-2S^{(2)}/3$. The three skewness
parameters have different weights of the bispectrum and hence the MFs can
extract more information on the configuration dependence of bispectrum
than using only one skewness value of $S$. The bispectrum
$B_{l_1l_2l_3}$ is defined as
\begin{equation}
\langle a_{l_1m_1}a_{l_2m_2}a_{l_3m_3}\rangle_c\equiv
\left(
\begin{tabular}{ccc}
$l_1$ & $l_2$ & $l_3$ \\
$m_1$ & $m_2$ & $m_3$
\end{tabular}
\right)
B_{l_1l_2l_3}.
\end{equation}
where $a_{lm}$ is the harmonic coefficients of a given temperature
anisotropy map.  The relation to the reduce bispectrum $b_{l_1l_2l_3}$
is $B_{l_1l_2l_3}=I_{l_1l_2l_3}b_{l_1l_2l_3}$.  We consider three
different types of NGs due to primordial bispectra: local type,
equilateral type, and orthogonal type. Single and multi-field
inflation models predict local-type NG (eq.[\ref{eq:local}]), which
generates the following form of CMB bispectrum \citep[e.g.,][]{KomatsuSpergel:01}:
\begin{equation}
\label{eq:bis_loc}
B_{l_1l_2l_3}=2f_{\rm NL}^{\rm (loc)}I_{l_1l_2l_3}
\int r^2dr[\alpha_{l_1}(r)\beta_{l_2}(r)\beta_{l_3}(r)+{\rm cyclic}],
\end{equation}
with
\begin{eqnarray}
\alpha_l(r)&\equiv &
\frac{2}{\pi}\int k^2dkg_{Tl}(k)j_l(kr), \\
\beta_l(r)&\equiv &
\frac{2}{\pi}\int k^2dkP_{\phi}(k)g_{Tl}(k)j_l(kr),
\end{eqnarray}
where $g_{Tl}$ is the radiation transfer function and $j_l$ is the
spherical Bessel function. We rewrite $f_{\rm NL}$ as $f_{\rm NL}^{\rm
  (loc)}$. The local-type NG is sensitive to the bispectrum with
squeezed configuration of triangle wavevectors ($l_1\ll l_2 \simeq
l_3$).

Other inflation scenarios with non-canonical kinetic terms
\citep{SeeryLidsey:05,ChenEasterLim:07}, Dirac-Born-Infeld models
\citep{Alishahihaetal:04}, and Ghost inflation
\citep{ArkaniHamedetal:04} predict large NG signals in equilateral
configuration triangles ($\ell_1\simeq \ell_2 \simeq
\ell_3$). Equilateral-type NG is characterized with $f_{\rm NL}^{\rm
  (eq)}$ defined as the amplitude of the following bispectrum
\citep{Babichetal:04}:
\begin{eqnarray}
\label{eq:bis_eq}
B_{l_1l_2l_3}^{\rm (eq)} = 6f_{\rm NL}^{\rm (eq)}I_{l_1l_2l_3}\int r^2 dr 
[-\beta_{l_1}(r)\beta_{l_2}(r)\alpha_{l_3}(r)~~~~~~~
\nonumber \\
 -\beta_{l_1}(r)\alpha_{l_2}(r)\beta_{l_3}(r)
-\alpha_{l_1}(r)\beta_{l_2}(r)\beta_{l_3}(r)
\nonumber \\
-2\delta_{l_1}(r)\delta_{l_2}(r)\delta_{l_3}(r)
+\{\beta_{l_1}(r)\gamma_{l_2}(r)\delta_{l_3}(r)+(5 {\rm permutation})\}],
\end{eqnarray}
where
\begin{eqnarray}
\gamma_l(r)&\equiv & \frac{2}{\pi}\int k^2dkP_{\phi}^{1/3}(k)g_{Tl}(k)j_l(kr), \\
\delta_l(r)&\equiv & \frac{2}{\pi}\int k^2dkP_{\phi}^{2/3}(k)g_{Tl}(k)j_l(kr).
\end{eqnarray}

The other type of NG which is sensitive to the bispectrum with a folded
triangle configuration ($l_1\simeq l_2 \simeq l_3/2$) is also considered by
\citet{SenatoreTassevZaldarriaga:09}.  This is called orthogonal-type NG and
characterized by $f_{\rm NL}^{\rm (ort)}$:
\begin{eqnarray}
\label{eq:bis_ort}
B_{l_1l_2l_3}^{\rm (ort)} = 6f_{\rm NL}^{\rm (ort)}I_{l_1l_2l_3}\int r^2 dr 
[-3\beta_{l_1}(r)\beta_{l_2}(r)\alpha_{l_3}(r)~~~~~~~ 
\nonumber \\
-3\beta_{l_1}(r)\alpha_{l_2}(r)\beta_{l_3}(r)-3\alpha_{l_1}(r)\beta_{l_2}(r)\beta_{l_3}(r) \nonumber \\
-8\delta_{l_1}(r)\delta_{l_2}(r)\delta_{l_3}(r) 
+3\{\beta_{l_1}(r)\gamma_{l_2}(r)\delta_{l_3}(r)+(5 {\rm perm.})\}].
\end{eqnarray}
As discussed in Sec. 4.2 of \citet{SenatoreTassevZaldarriaga:09}, we
include the integration of $r$ further than the last scattering
surface $r_\star$ to calculate the equilateral and orthogonal-type
bispectra. This reduces the sensitivity of these types of NG.

We also take into account the effect of unmasked point sources (e.g.,
radio galaxies) which generates an additional NG in observed CMB
maps. Assuming them to be Poisson distribution, the bispectrum has a
constant value for all configurations of wavevectors:
\begin{equation}
B_{l_1l_2l_3}^{\rm (ps)}=b^{\rm (ps)}I_{l_1l_2l_3}. 
\end{equation}
where $b^{\rm (ps)}$ is a constant value.

\begin{figure*}
\caption{Three Skewness values (eq.[\ref{eq:skewness}]) for local-type
  (eq.[\ref{eq:bis_loc}]), equilateral-type (eq.[\ref{eq:bis_eq}]), and
  orthogonal-type (eq.[\ref{eq:bis_ort}]) NGs and point sources 
  as a function of the Gaussian smoothing scale $\theta$. We add WMAP beam
  functions for V+W co-added maps and a pixel window function
  corresponding to each $\theta$ listed in Table \ref{tab:scale}.}
\begin{center}
\includegraphics[width=4.5cm]{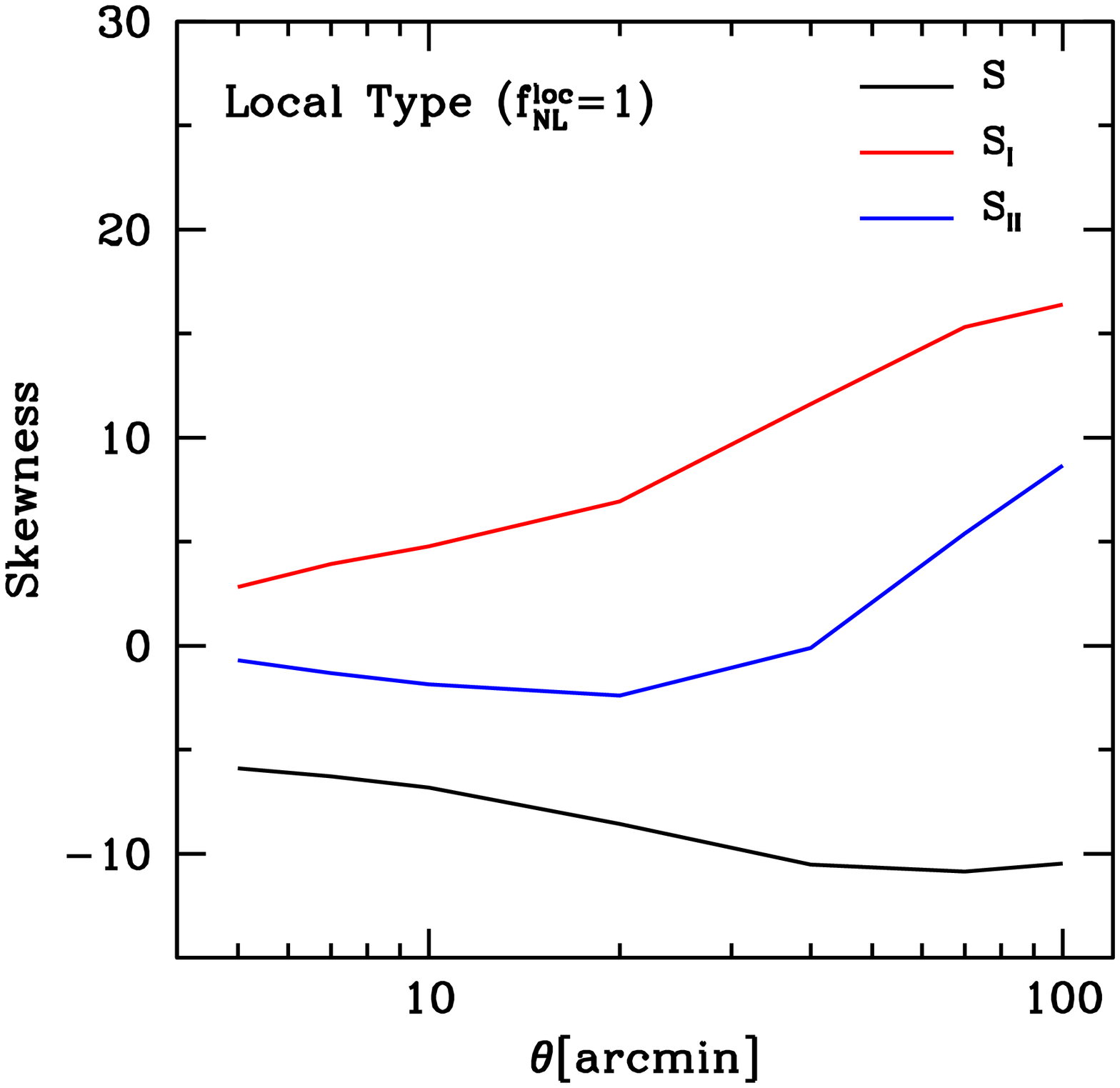}
\includegraphics[width=4.5cm]{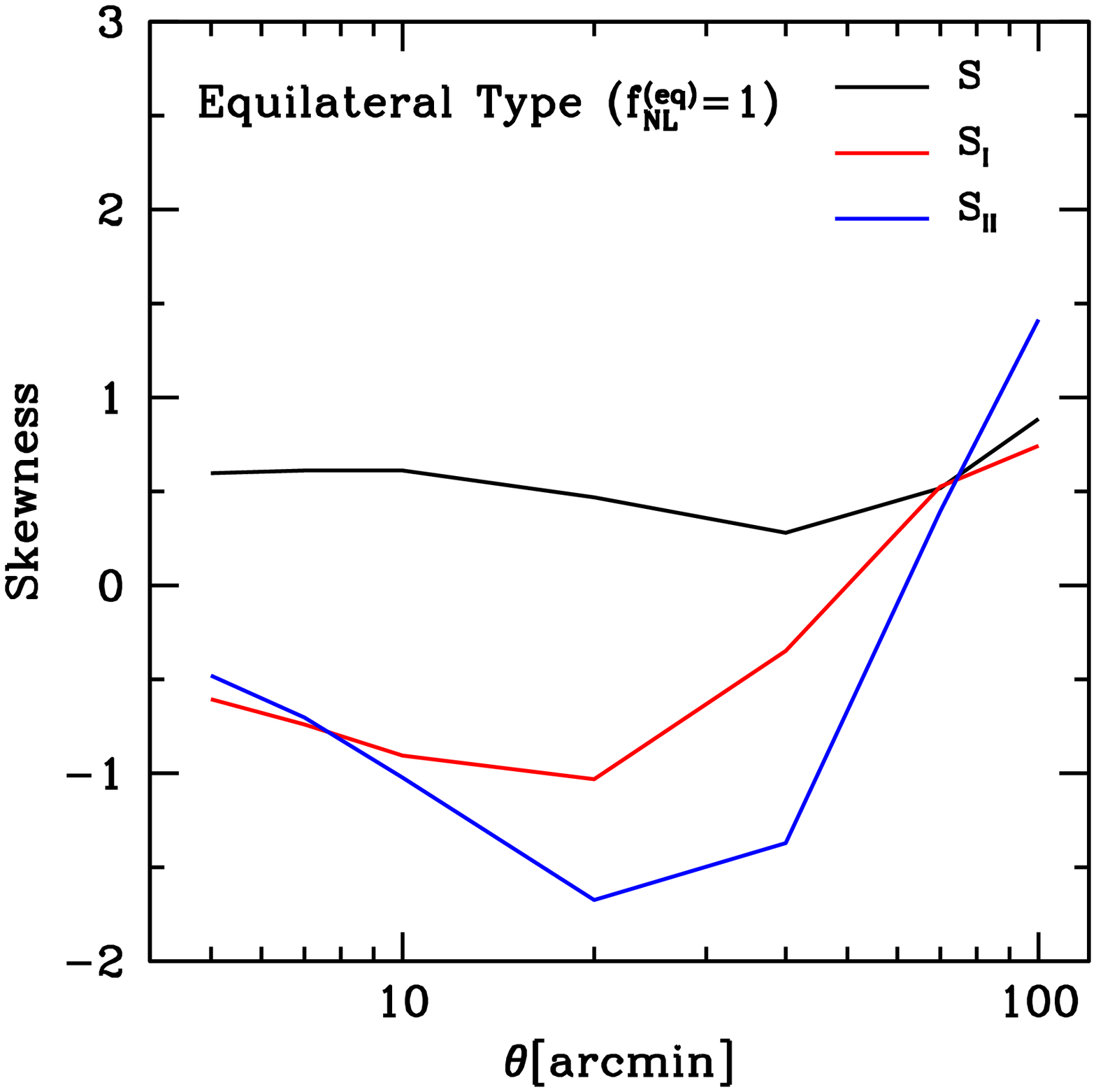}
\includegraphics[width=4.5cm]{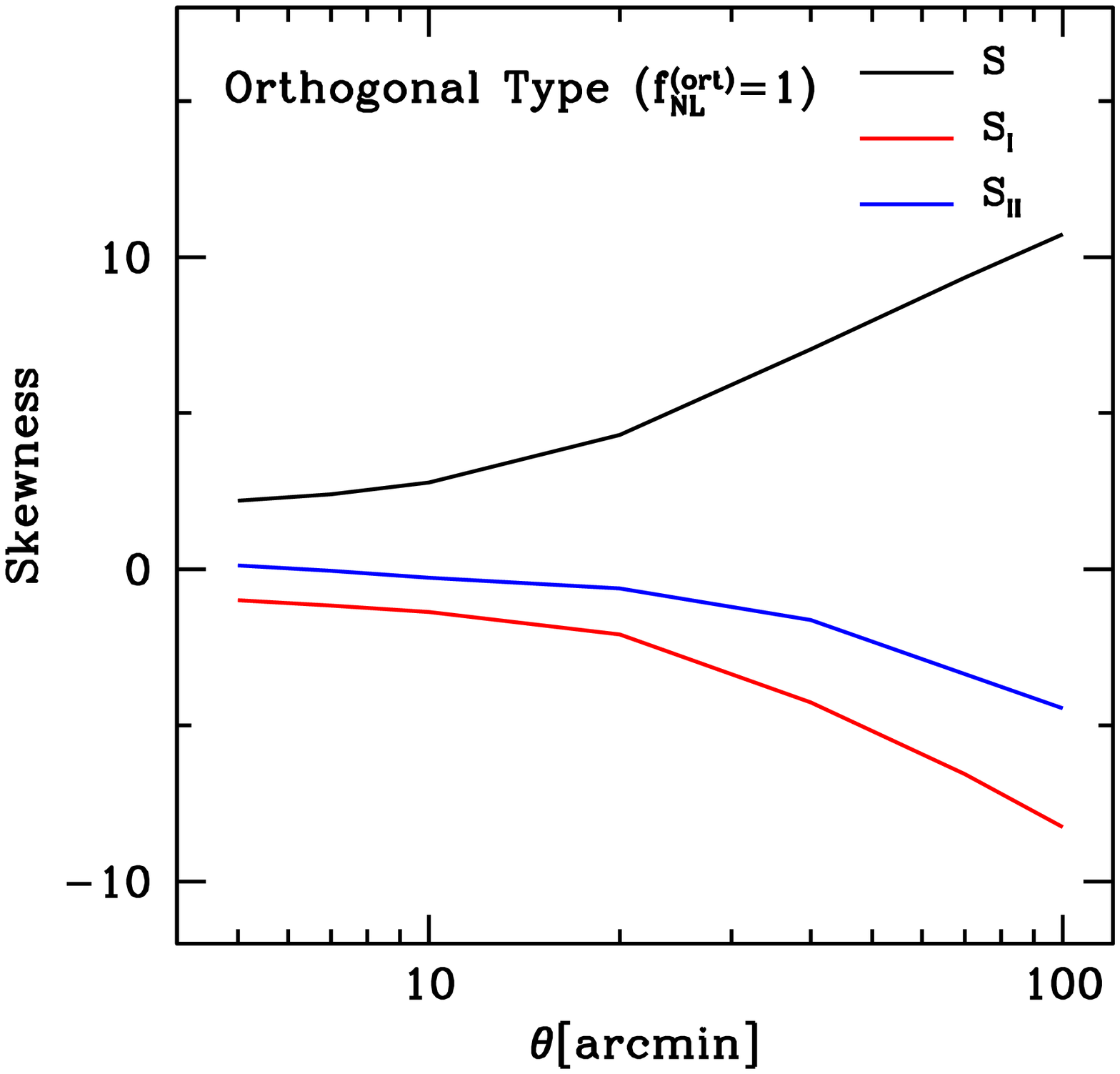}
\includegraphics[width=4.5cm]{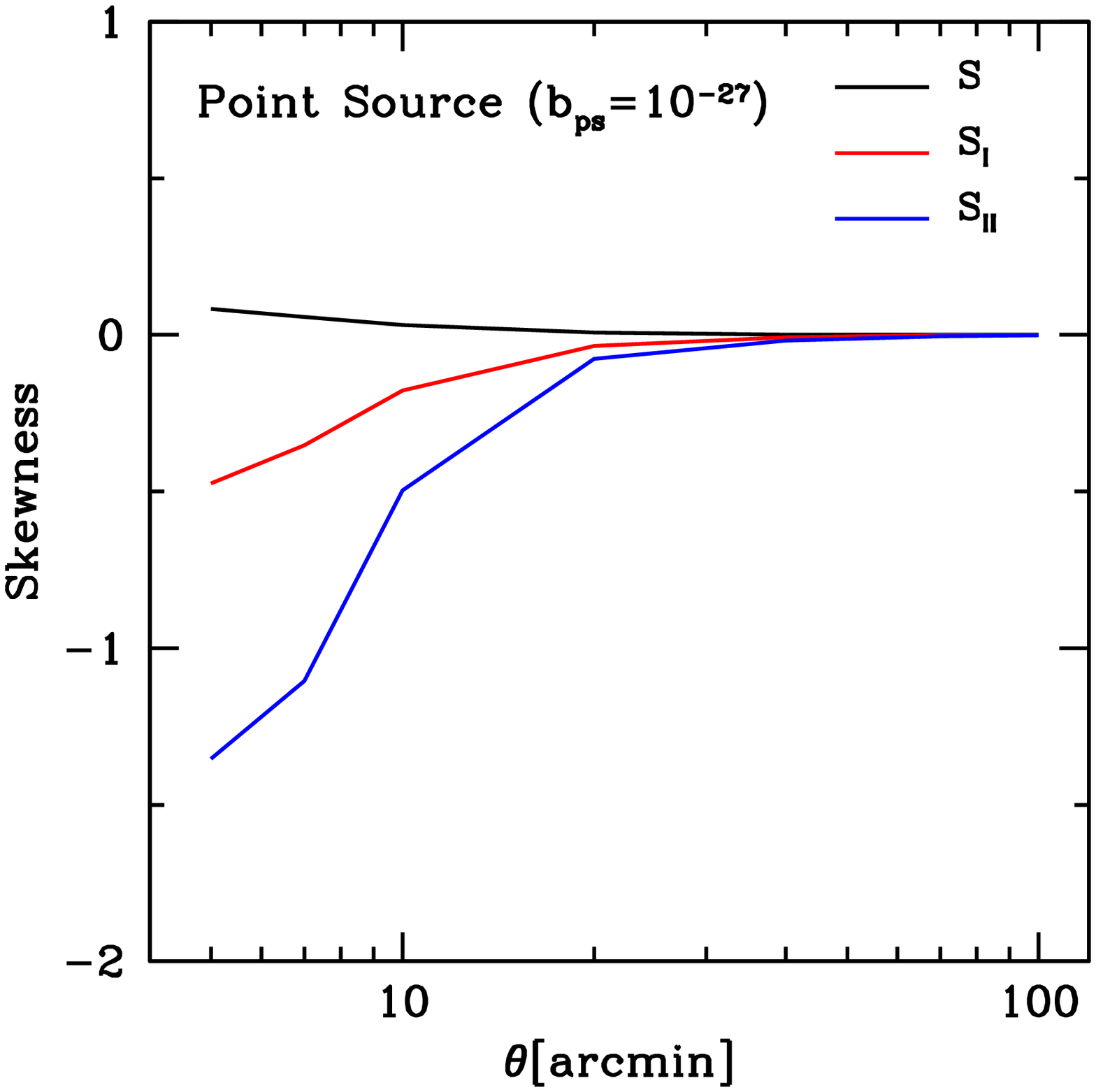}
\label{fig:skew}
\end{center}
\end{figure*}

Fig. \ref{fig:skew} plots the smoothing scale dependence of three
skewness parameters (eq.[\ref{eq:skewness}]) from the local-type,
equilateral-type, orthogonal-type primordial NG components with the
unity value of $f_{\rm NL}$ and from unmasked point source NG with the
constant bispectrum $b^{\rm (ps)}=10^{-27}$. We add noise and beam
functions of WMAP 7-year V+W co-added maps which become important at
smaller smoothing scale such as $\theta<10$ arcmin. Scale dependences
of skewness parameters are quite different among different NG types.
Measuring MFs of CMB maps with different smoothing scales is important
to break degeneracy of different NG sources.

\subsubsection{Second-order perturbation}
\citet{Matsubara:10} has derived the second-order corrections of MFs,
which are characterized with the product of skewness parameters and four
kurtosis parameters:
\begin{eqnarray}
\label{eq:mf_pb2}
v_0^{(2)}(\nu)=\frac{S^2}{72}H_5(\nu) + \frac{K}{24}H_3(\nu),~~~~~~~~~~~~~~~~~~~~~~~~~~~~~~ \\
v_1^{(2)}(\nu)=\frac{S^2}{72}H_6(\nu) + \frac{K-SS_{\rm I}}{24}H_4(\nu)~~~~~~~~~~~~~~~~~~~~~~~ \nonumber \\
-\frac{1}{12}\left(K_{\rm I}+\frac38S_{\rm I}^2\right)H_2(\nu)-\frac{K_{\rm III}}{8},~~~~~~~~~~~~~~~~~~~ \\
v_2^{(2)}(\nu) = \frac{S^2}{72}H_7(\nu)+\frac{K-2SS_{\rm I}}{24}H_5(\nu)~~~~~~~~~~~~~~~~~~~~~~ \nonumber \\
-\frac16\left(K_{\rm I}+\frac12SS_{\rm II}\right)H_3(\nu) -\frac12\left(K_{\rm II}+\frac12S_{\rm I}S_{\rm II}\right)H_1(\nu).
\end{eqnarray}
The kurtosis parameters are defined as
\begin{eqnarray}
\label{eq:kurtosis}
K\equiv\frac{\langle f^4\rangle_c}{\sigma_0^4},~~~~~~~~~~~~~~~~~~~~~~~~~~~~~~~
K_{\rm I}\equiv \frac{\langle (\nabla^2f)f^3\rangle_c}{\sigma_0^4\sigma_1^2}, \\ 
K_{\rm II}\equiv\frac{2\langle f|\nabla f|^2\nabla^2 f\rangle_c+\langle|\nabla f|^4\rangle_c}{\sigma_0^2\sigma_1^4},~~~~
K_{\rm III}\equiv\frac{\langle|\nabla f|^4\rangle_c}{2\sigma_0^2\sigma_1^4},
\end{eqnarray}
where $\langle ...\rangle_c$ denotes the ensemble average of the
connected part. 

For the local-type NG, the reduced trispectrum is written as 
\begin{eqnarray}
\label{eq:tri_full}
{\cal T}^{l_1l_2}_{l_3l_4}(L)=I_{l_1l_2L}I_{l_3l_4L}
\left\{\frac{25}{9}\tau_{\rm NL}\int r_1^2dr_1r_2^2dr_2F_L(r_1,r_2)~~\right. \nonumber \\
\times \alpha_{l_1}(r_1)\beta_{l_2}(r_1)\alpha_{l_3}(r_2)\beta_{l_4}(r_2) \nonumber \\
\left. + g_{\rm NL}\int r^2dr\beta_{l_2}(r)\beta_{l_4}(r) 
[\alpha_{l_1}(r)\beta_{l_3}(r)+\beta_{l_1}(r)\alpha_{l_3}(r)]\right\},
\end{eqnarray}
where
\begin{equation}
F_L(r_1,r_2)\equiv \frac{2}{\pi}\int k^2dkP_\phi(k)j_L(kr_1)j_L(kr_2).
\end{equation}
For the single-field inflation model with local-type NG
(eq.[\ref{eq:local}]), $\tau_{\rm NL}$ is equal to $36f_{\rm NL}^{\rm
  (loc) 2}/25$.

\begin{figure}
\caption{Trispectrum of two-dimensional CMB maps is calculated at each
  configuration of tetrahedron. A tetrahedron is uniquely determined by
  three side lengths $l_1, l_2, l_3$ and their open angles
  $\theta_{12}$ and $\theta_{23}$. The other side length $l_4$ and the
  diagonal lines $l_{12}$ and $l_{23}$ used in the equation
  (\ref{eq:tri_flat}) are also shown.}
\begin{center}
\includegraphics[width=6cm]{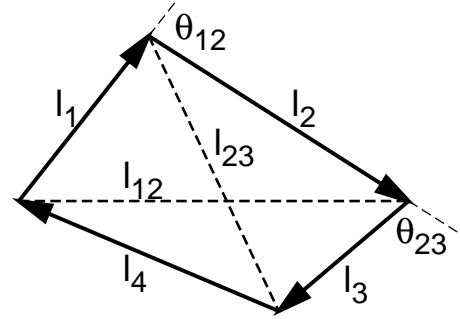}
\label{fig:tetrahedron}
\end{center}
\end{figure}

The kurtosis parameters are given by a summation over all
configurations of trispectra. The full calculation of trispectra is,
however, computationally very expensive and hence we estimate kurtosis values
using Monte-Carlo integration of the flat-sky approximation.
\citet{Matsubara:10} finds that the full-sky spectrum with its
multipole configuration of $\{l_i\}$ is well approximated by the flat-sky
spectrum with the wavelength configuration of ${l_i+1/2}$:
\begin{eqnarray}
\label{eq:flatskyapprox}
{\cal T}^{l_1l_2}_{l_3l_4}(L)\simeq  I_{l_1l_2L}I_{l_3l_4L}~~~~~~~~~~~~~~~~~~~~~~~~~~~~~~~~~~~~ 
\nonumber \\
\times {\cal T}\left(l_1+\frac12,l_2+\frac12,l_3+\frac12,l_4+\frac12;L+\frac12\right).
\end{eqnarray}
The proportional factor $I_{l_1l_2L}I_{l_3l_4L}$ has non-zero value
when both $l_1+l_2+L$ and $l_3+l_4+L$ have even number.  Note that the
difference of 1/2 for all arguments of lengths $l$ is very important
for the accurate estimation. The kurtosis in flat-sky approximation is
given as
\begin{eqnarray}
\label{eq:kurtosis_flat}
K_A = \frac{1}{\sigma_0^6}\int\frac{l_1dl_1}{2\pi}\frac{l_2dl_2}{2\pi}\frac{l_3dl_3}{2\pi}
\frac{d\theta_{12}}{2\pi}\frac{d\theta_{23}}{2\pi}\tilde{K}_A(l_1,l_3,l_{12}) \nonumber \\
\times T(l_1,l_2,l_3,l_4;l_{12},l_{23})W(l_1)W(l_2)W(l_3)W(l_4),
\end{eqnarray}
where 
\begin{eqnarray}
\tilde{K}=1,~~~~\tilde{K}_{\rm I}=-\frac{l_1^2}{2q^2},
~~~~~\tilde{K}_{\rm II}=-\frac{l_{12}^2-4l_1^2l_3^2}{16q^4}, \nonumber  \\
\tilde{K}_{\rm III}=\frac{l_{12}^4+4l_1^2(l_3^2-l_{12}^2)}{32q^4},
\end{eqnarray}
and the Gaussian window function in the flat approximation is now
given by $W(l)=\exp[-\{(l+1/2)\theta\}^2/2]$.  The trispectrum $T$ is
symmetric against the arbitrary exchange of arguments among $l_1$,
$l_2$, $l_3$ and $l_4$ and then it is given as a sum of the reduced
trispectrum ${\cal T}$:
\begin{eqnarray}
\label{eq:tri_flat}
T(l_1,l_2,l_3,l_4;l_{12},l_{23}) =
P(l_1,l_2,l_3,l_4,l_{12})~~~~~~~~~~~~~~ \nonumber \\ +
P(l_1,l_3,l_2,l_4,l_{13}) + P(l_1,l_4,l_2,l_3,l_{23}),
\end{eqnarray}
and 
\begin{eqnarray}
P(l_1,l_2,l_3,l_4,L)={\cal T}(l_1,l_2,l_3,l_4,L) + {\cal T}(l_2,l_1,l_3,l_4,L)~~~~ \nonumber \\
+ {\cal T}(l_1,l_2,l_4,l_3,L) + {\cal T}(l_2,l_1,l_4,l_3,L).
\end{eqnarray}
As shown in Fig. \ref{fig:tetrahedron}, the quadrangle configuration
is uniquely determined by five parameter spaces: three side lengths
$l_1,l_2,l_3$ and their open angles $\theta_{12}$, $\theta_{23}$. The
other side length $l_4$, two diagonal lengths $l_{12}$ and $l_{23}$
(see Fig. \ref{fig:tetrahedron}) and
$l_{13}=|\mathbf{l_1}+\mathbf{l_3}|$ are written in terms of the five
parameters. Fortunately, the configuration dependence of the
trispectrum is smooth for the local-type NG model and hence Monte
Carlo integration is applicable to estimate kurtosis values in shorter
time. The integration of the side lengths $l_1, l_2$ and $l_3$ is done
from 2 to $l_{\rm max}$ given by Table \ref{tab:scale} and that of
$\theta_{12}$ and $\theta_{23}$ is done from 0 to $2\pi$. In equation
(\ref{eq:flatskyapprox}), the fractional values of the length
$\{l_i\}$ are allowed in the flat-sky approximation of ${\cal T}$,
while the full-sky trispectrum ${\cal T}^{l_1l_2}_{l_3l_4}(L)$ is only
given for a set of integer values $l_i$. We use the round-off values
of $\{l_i\}$ to get ${\cal T}$. We get the reasonable values of
kurtosis parameters by ${\cal O}(10^9)$ calculation, which takes much
shorter time than the full calculation of the trispectrum.

For the trispectrum, we also consider the effect of point source as
\begin{equation}
{\cal T}^{l_1l_2}_{l_3l_4}(L)=t^{\rm (ps)}I_{l_1l_2L}I_{l_3l_4L},
\end{equation}
where $t^{\rm (ps)}$ is a constant value.

\begin{table}
\caption{Resolution scale $N_{\rm side}$ for HEALPix (total number of
  pixels is given by $12N_{\rm side}^2$) and the maximum values of
  multipole $l_{\rm max}$ at each smoothing scale $\theta$ [arcmin].}
\begin{center}
\begin{tabular}{ccc}
\hline\hline
$\theta$ [arcmin] & $N_{\rm side}$ & $l_{\rm max}$ \\
\hline
 100  &  128  &  136 \\
 70   &  128  &  196 \\
 40   &  256  &  340 \\
 20   &  512  &  684 \\
 10   &  512  &  750 \\
 7    &  512  & 1024 \\
 5    &  512  & 1250 \\
\hline
\end{tabular}
\label{tab:scale}
\end{center}
\end{table}

\section{Comparison of Perturbation Theory with Numerical Simulations}
\label{sec:simulations}

\subsection{Computation of Minkowski Functionals from CMB maps}

Computation method of the MFs of CMB maps are described in Appendix
A.1. of \citet{Hikageetal:06}. The range of $\nu$ is from $-3.6$ to
$3.6$ and the binning number is set to be 36 for each MF with the
equal binning width $\Delta\nu=0.2$.  The numerical estimations of 2nd
and 3rd MFs do not completely agree with the analytical predictions
even in Gaussian fields. \citet{LimSimon:12} found that the deviation
comes from the approximation of the delta function with a finite
difference. In Gaussian fields, the correction terms of the finite
binning effect of the 2nd and 3rd MFs are given by
\begin{eqnarray}
R_k(\nu)&\equiv&\left[\frac{1}{\Delta\nu}\int^{\nu+\Delta\nu/2}_{\nu-\Delta\nu/2}dxV_k(x)\right]-V_k^{\rm (G)}(\nu) \nonumber \\
&=&\frac{(\Delta\nu)^2}{24}H_{k+1}(\nu)A_ke^{-\nu^2/2}+{\cal O}(\Delta\nu^4),
\end{eqnarray}
where $\Delta\nu$ is the binning width of $\nu$.  We subtract the
correction terms from the measured MFs.

\subsection{Comparison of Perturbation Theory with Non-Gaussian CMB Maps}
Actual CMB maps have various observational effects such as survey
geometry, inhomogeneous noise, which may cause systematic uncertainty
in the NG measurements from MFs. We take into account such
observational systematics by constructing realistic CMB simulation
maps and test if the perturbation works for actual observations.
\subsubsection{Full Non-Gaussian Simulations}
\label{sssec:full}
We employ 1000 realizations of simulated CMB maps with a local-type NG
available in public \citep{ElsnerWandelt:09}. These simulation maps
include full radiation transfer function. We make mock CMB maps as
described in \citet{Hikageetal:08}: making Gaussian CMB maps with
their input power spectra following WMAP 7-year cosmology and
including the beam function for each differencing assembly (DA), we
add Gaussian-distributed noise to each pixel with the standard
deviation of $\sigma/N_{\rm obs}$ in WMAP 7-year observations. We
co-add noise-included DA maps with the inverse weight of averaged
noise variance and then mask the area outside the KQ75 mask of Galactic
foregrounds and point sources \citep{Goldetal:11}.

\begin{figure*}
\caption{{\it Left:} variance, skewness and kurtosis of NG simulations
  with $f_{\rm NL}^{\rm (loc)}=300$ using CMB simulation maps with
  local-type NG \citep{ElsnerWandelt:09} (symbols). The lines show the
  theoretical predictions based on the perturbation theory. {\it
    Right:} second-order correction of MFs in the same NG simulation
  maps smoothed at $\theta=10$ arcmin (symbols) by subtracting the
  Gaussian term $V_k^{\rm (G)}$ and the first-order correction
  $V_k^{\rm (1)}$.  The NG simulations do not include the $g_{\rm NL}$
  component and thereby the second-order correction comes from
  $\tau_{\rm NL}$ corresponding to $36f_{\rm NL}^{\rm (loc)
    2}/25\simeq 1.296\times 10^5$. For comparison, perturbative
  predictions are written in the lines. The simulations include the
  WMAP beam and noise for the V+W co-added map and a pixel window
  function at each $\theta$.}
\begin{center}
\includegraphics[width=8cm]{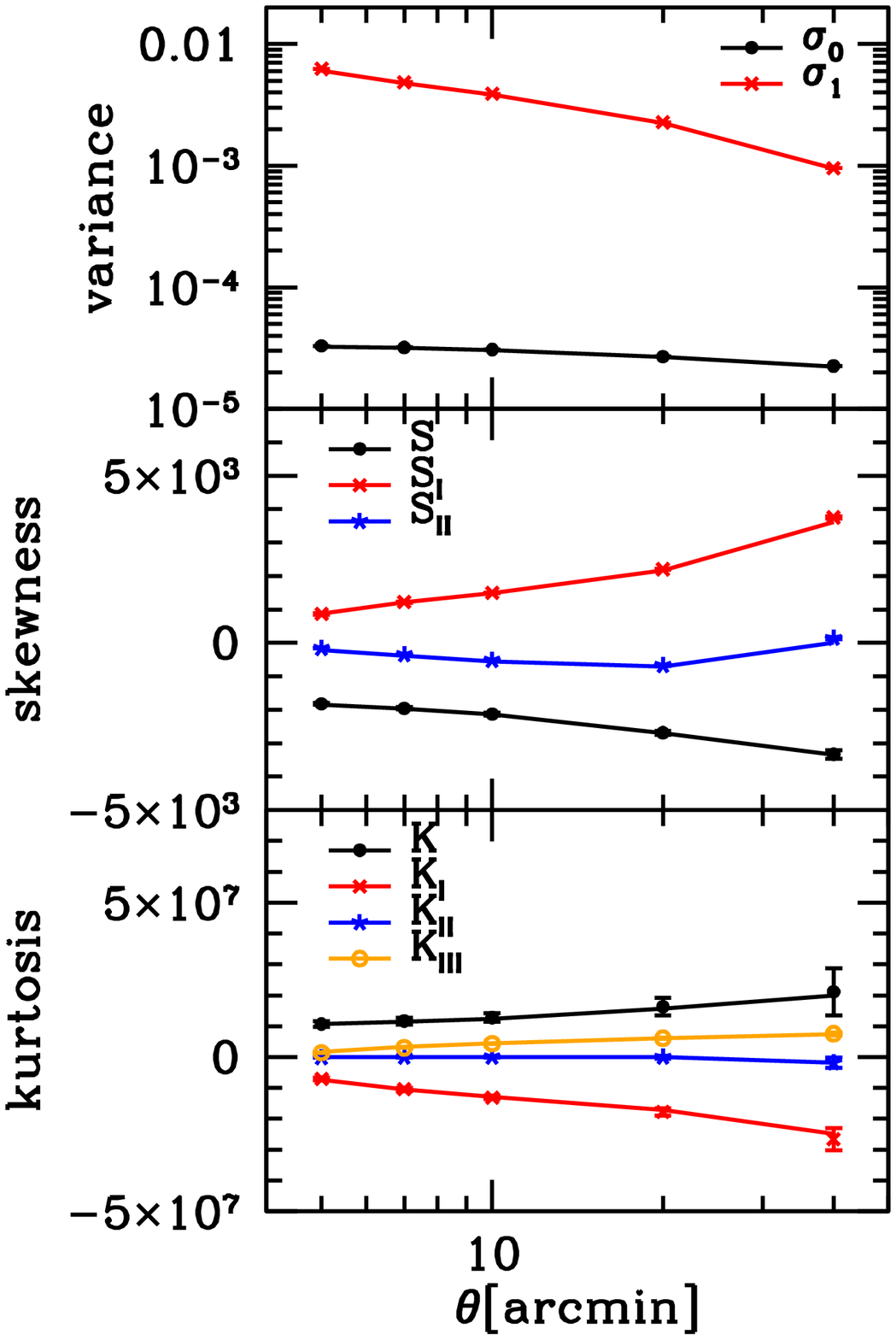}
\includegraphics[width=8cm]{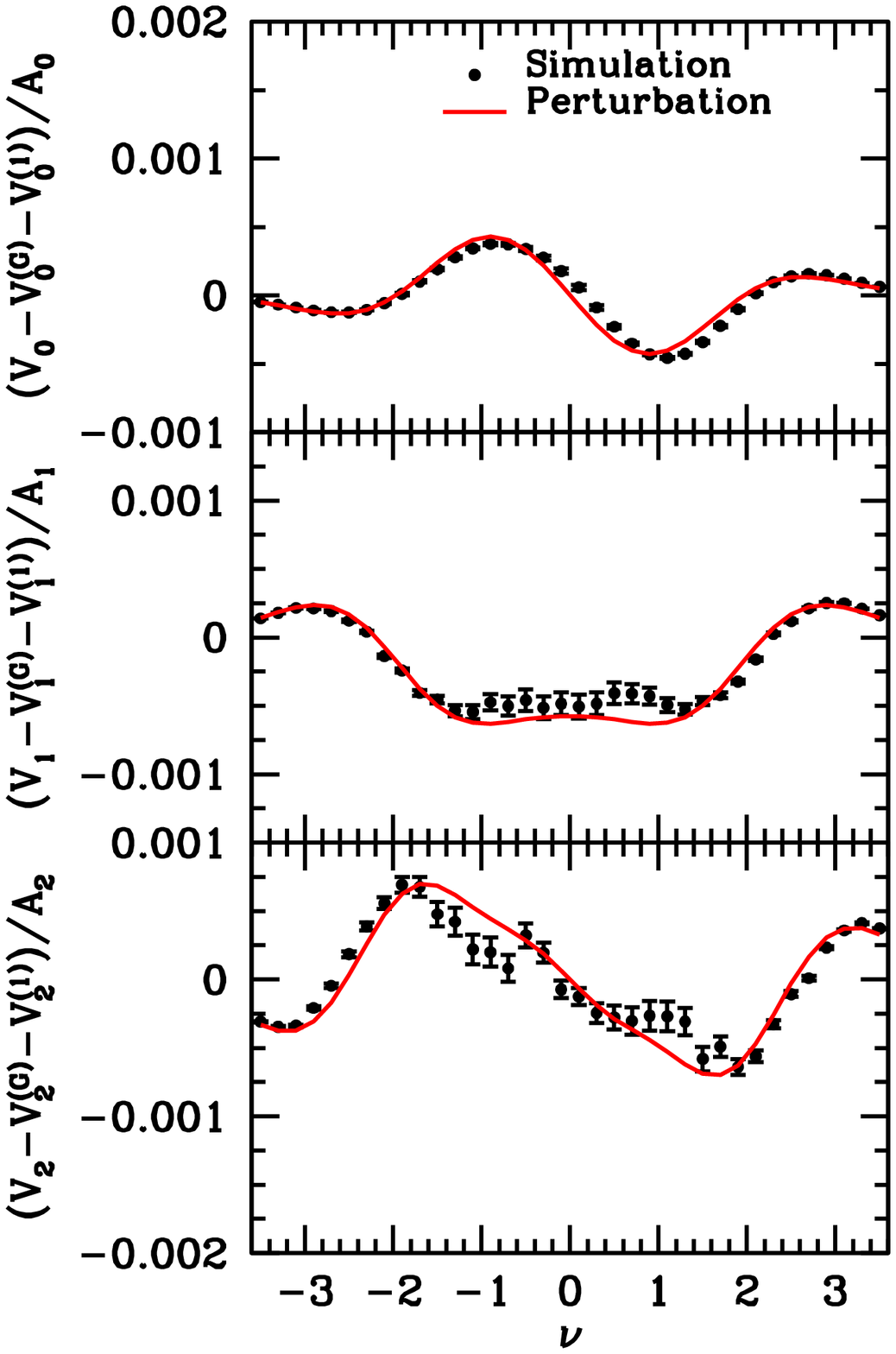}
\label{fig:mffull}
\end{center}
\end{figure*}

Fig. \ref{fig:mffull} (left-hand panels) shows a comparison of two
variances (eq.[\ref{eq:var}]), three skewness parameters
(eq.[\ref{eq:skewness}]), and four kurtosis parameters
(eq.[\ref{eq:kurtosis}]) with $f_{\rm NL}^{\rm (loc)}=300$ between
analytical estimations and simulation results at different values of
$\theta$s.  The simulations do not include the $g_{\rm NL}$ term and
thereby the second-order NG comes from the square of the $f_{\rm NL}$
term, or $\tau_{\rm NL}$. The error-bars represent the 1$\sigma$
dispersion of simulation results divided by the square root of 1000,
that is, the number of realizations.  The right-hand panels show that the
2nd-order correction of each MF at $\theta=10$ arcmin. In the plot of
MFs, we subtract Gaussian and 1st-order perturbative correction to
focus on the 2nd-order correction (\citet{Hikageetal:08} have already
shown that the 1st-order perturbative correction of MFs due to $f_{\rm
  NL}^{\rm (loc)}$ agree with the results from NG simulations). We
find that the 2nd-order perturbative corrections of MFs also agree
with simulation results very well even including the observational
effects.

\subsubsection{Sachs-Wolfe approximations}
\label{sssec:sw}
\begin{figure*}
\caption{{\it Left:} same as Fig. \ref{fig:mffull} but for NG
  simulations in the Sachs-Wolfe approximation. The NG parameters are
  $f_{\rm NL}^{\rm (loc)}=100$ and $g_{\rm NL}=10^6$ in which the
  contribution of $\tau_{\rm NL}$ is negligible. {\it Right:}
  comparison of the NG correction of each MF between the perturbative
  theory (solid lines) and the NG simulations (filled circles). First-
  and second-order perturbative corrections are shown with the dotted
  and dashed lines, respectively.}
\begin{center}
\includegraphics[width=8cm]{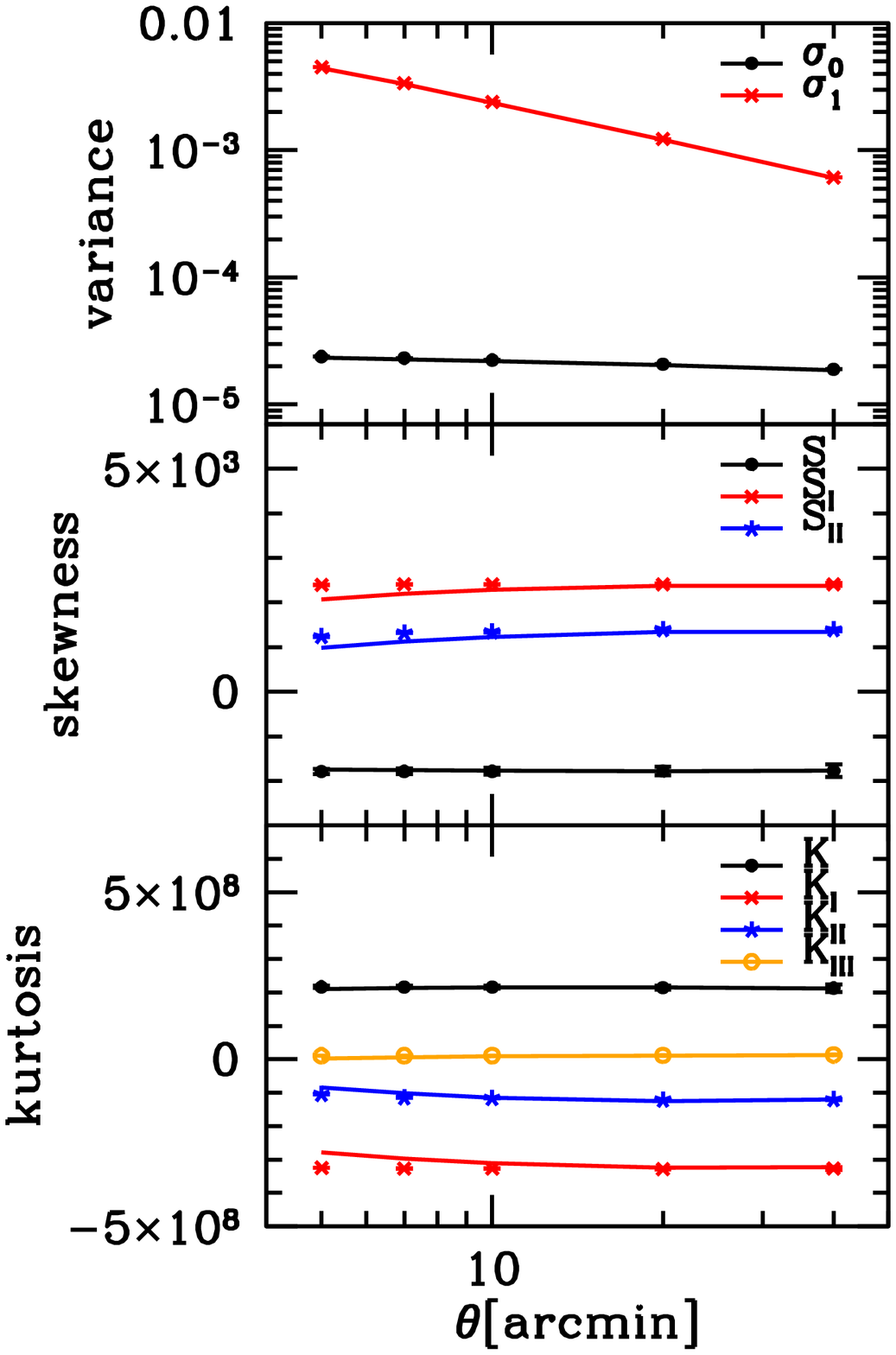}
\includegraphics[width=8cm]{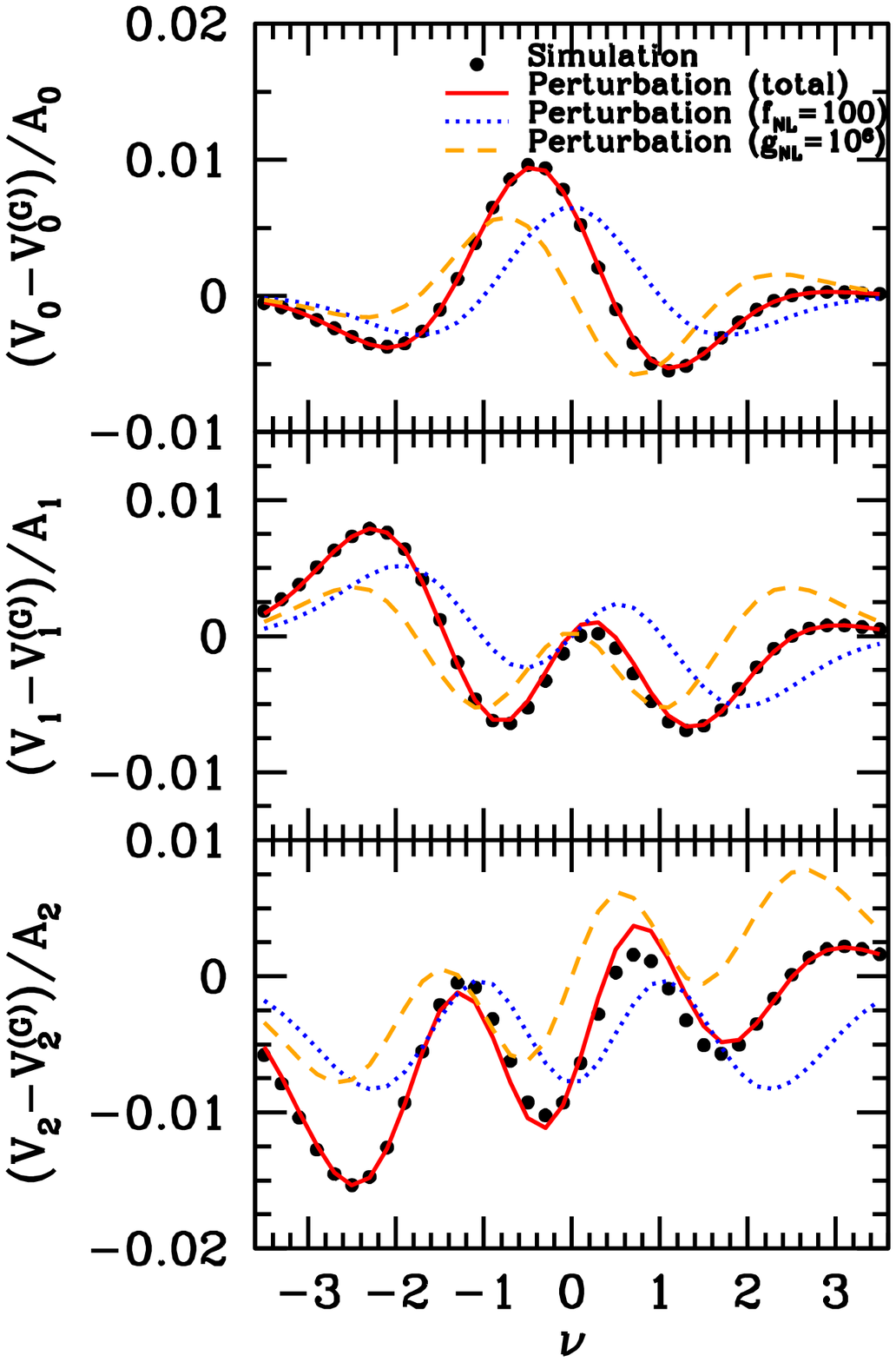}
\label{fig:mfsw}
\end{center}
\end{figure*}

We also compare the 2nd-order correction due to $g_{\rm NL}$ type
using NG simulation maps in Sachs-Wolfe approximation where the NG is
locally given by
\begin{equation}
\frac{\Delta T_{\rm SW}}{T}=\frac{\Delta T_{\rm G}}{T}-3f_{\rm NL}^{\rm
(loc)}\left(\frac{\Delta T_{\rm G}}{T}\right)^2 +9g_{\rm
NL}\left(\frac{\Delta T_{\rm G}}{T}\right)^3.
\end{equation}
The angular power spectrum is set to be $l(l+1)C_l^{\rm
  SW}/2\pi=10^{-10}$ at $l<=l_{\rm max}$ where $l_{\rm max}$ is given
in Table \ref{tab:scale} at each $\theta$.  We generate 6000
realizations with $f_{\rm NL}^{\rm (loc)}=100$ and $g_{\rm
  NL}=10^6$. For these values, the 1st-order correction from $f_{\rm
  NL}^{\rm (loc)}$ and the 2nd-order correction from $g_{\rm NL}$ have
comparable amplitudes, but the contribution of $\tau_{\rm NL}\sim
f_{\rm NL}^{\rm (loc) 2}$ is negligible. Fig. \ref{fig:mfsw} shows a
similar plot to that of Fig. \ref{fig:mffull} but for the comparison
with simulations in the Sachs-Wolfe approximation. The right-hand
panel shows the MFs subtracting only Gaussian terms at $\theta=10$
arcmin. \citet{Matsubara:10} showed that the perturbation formulae
work very well in the Sachs-Wolfe approximation. Here we find that the
perturbation also works even including the various observational
effects such as survey mask, inhomogeneous noise and beam window
function. The excellent agreement indicates that the flat-sky
approximation and Monte Carlo integration also work very well at a
wide range of scales.

\subsection{Distribution of NG parameters estimated from simulations}
We estimate the bestfit values and the errors of each NG parameter by
the least chi-square fitting of observed MFs with the perturbative
predictions:
\begin{equation}
\label{eq:chisq}
\chi^2=\sum_{i,j}[V_i^{\rm (obs)} - V_i^{\rm (theory)}(p_k)]C_{ij}^{-1}[V_j^{\rm (obs)} - V_j^{\rm (theory)}(p_k)],
\end{equation}
where $p_k$ denote NG parameters such as $f_{\rm NL}$s, $\tau_{\rm
  NL}$ and $g_{\rm NL}$ and subscript $i$ and $j$ of MFs denote
different bins of threshold $\nu$, different kinds of MFs, and
different smoothing scales $\theta_s$. We estimate the covariance
matrix of MFs using 6000 realizations of Gaussian CMB maps including
WMAP7 observational effects described in the previous section. For the
chi-square measurement, we reduce the binning number to 18 from 36
original bins.  The reduction of binning does not affect the results,
which means that the binning number of $18$ is enough that the result
is converged.  The maximum number of the side length of the covariance
matrix is 270 corresponding to 18 bins of $\nu$ $\times$ three kinds
of MFs $\times$ five different $\theta$. We have checked that 6000
realizations are enough for the fitting results to be converged.

\begin{figure*}
\caption{{\it Left:} distribution of the bestfit values of $f_{\rm
    NL}^{\rm (loc)}$ estimated from 1000 NG simulation maps in which
  the input values of $f_{\rm NL}^{\rm (loc)}=\pm 100$; {\it Middle:}
  distribution of $\tau_{\rm NL}$ estimated from 1000 NG simulation
  maps with the input values of $f_{\rm NL}^{\rm (loc)}=\pm 300$, that
  is, the corresponding $\tau_{\rm NL}=1.296\times 10^5$; {\it Right:}
  distribution of $g_{\rm NL}$ estimated from 1000 NG simulations in
  the Sachs-Wolfe approximations.  The input values of NG parameters
  are $f_{\rm NL}^{\rm (loc)}=100$ and $g_{\rm NL}=10^6$. The lines
  show the Gaussian distribution with its mean of the input values and
  the dispersion given by the square root of the inverse of the Fisher
  matrix $(F^{-1})_{kk}^{1/2}$.}
\begin{center}
\includegraphics[width=6cm]{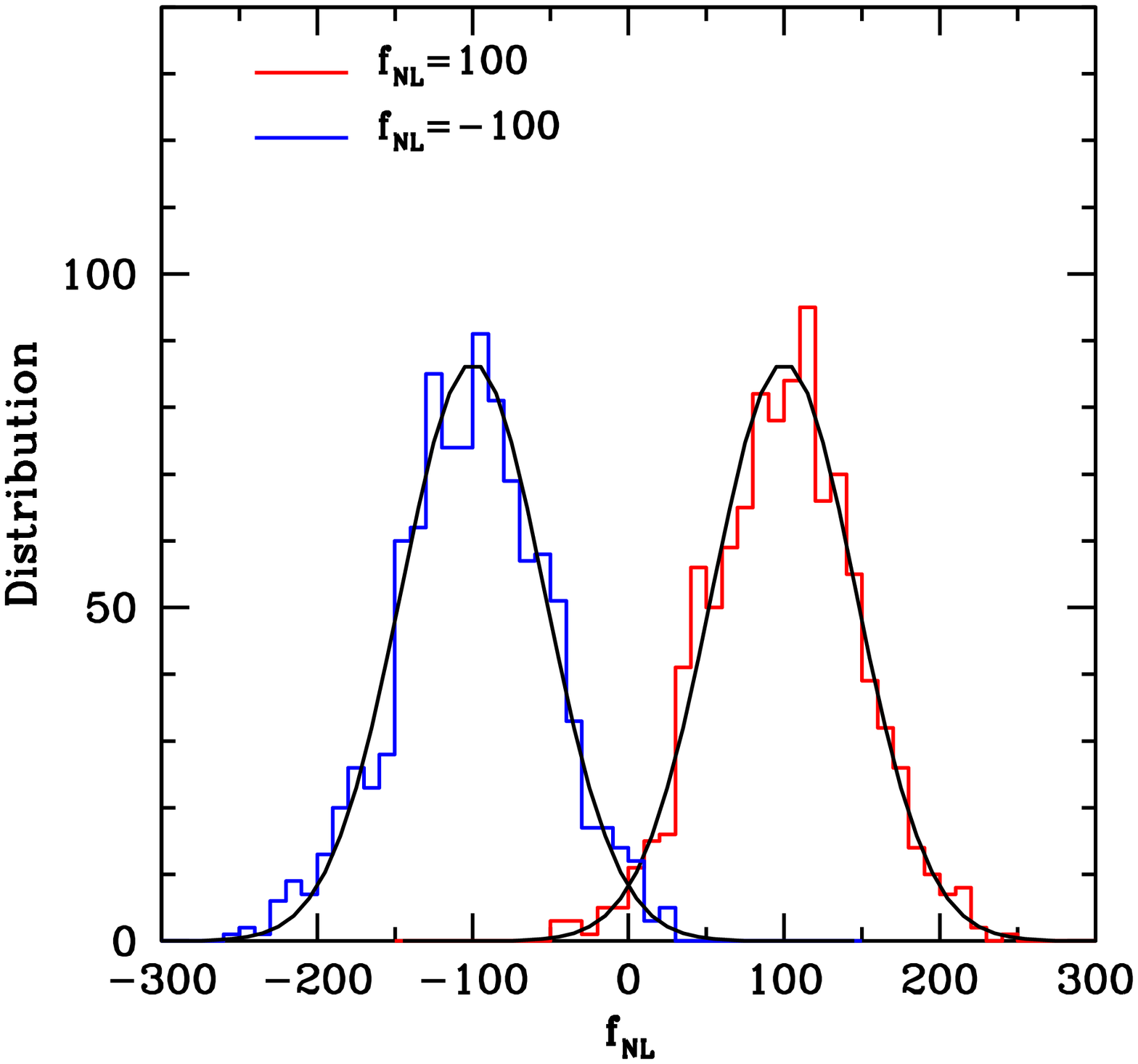}
\includegraphics[width=5.7cm]{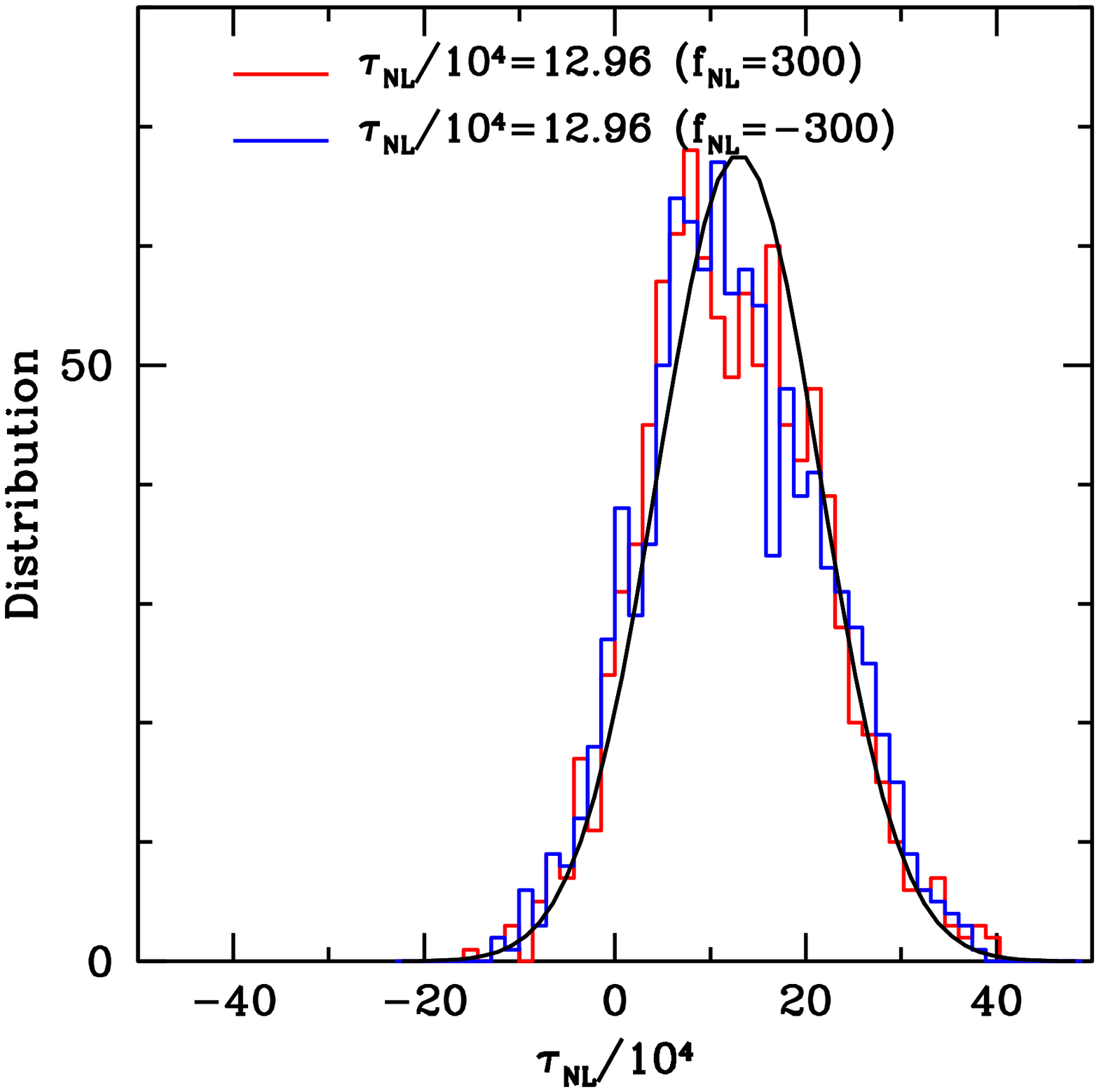}
\includegraphics[width=6cm]{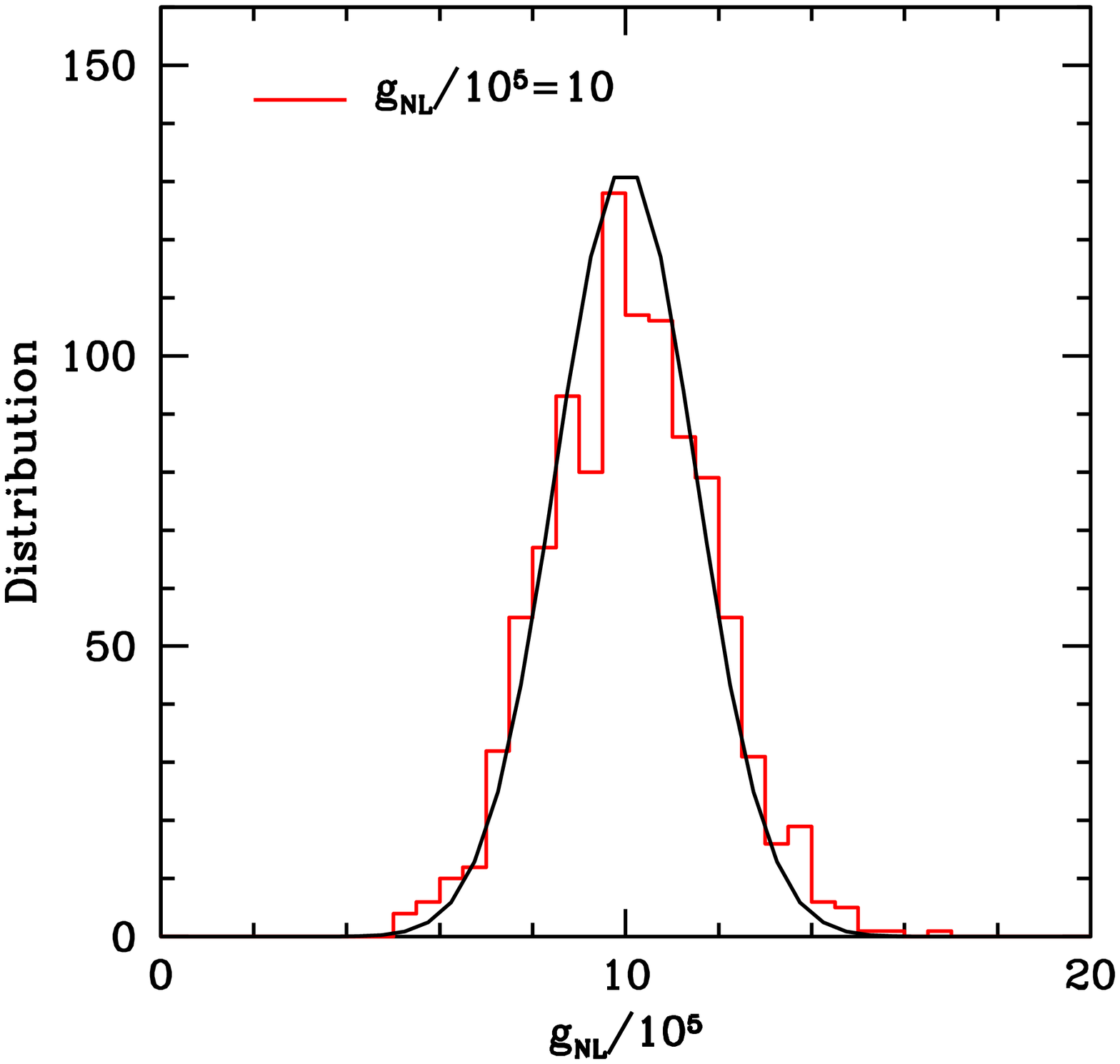}
\label{fig:dist_ftgnl}
\end{center}
\end{figure*}

Assuming that the covariance derivative $\partial C_{ij}/\partial p_i$
is negligible, the Fisher matrix is simply given by
\begin{equation}
\label{eq:fisher}
F_{kk'}=\sum_{ij} \frac{\partial V_i}{\partial
  p_k}C_{kk'}^{-1}\frac{\partial V_j}{\partial p_{k'}}.
\end{equation}
We use NG CMB simulation maps and test if the above chi-square
estimations generate the expected distribution of $k$-th NG parameter
with the mean of the input value and the error expected from the
Fisher matrix as $[(F^{-1})_{kk}]^{1/2}$. Fig. \ref{fig:dist_ftgnl}
shows the distribution of best-fit values of $f_{\rm NL}$, $\tau_{\rm
  NL}$ from full NG simulations and $g_{\rm NL}$ from NG simulations
in Sachs-Wolfe approximation. The details of NG simulations are
written in the previous subsection. Averaged values are respectively
$f_{\rm NL}= 100\pm 47 (-100\pm 48)$ against the input of $100(-100)
\pm 48$, and $\tau_{\rm NL}/10^4=12.3 \pm 8.6 (12.4 \pm 9.1)$ against
the input of $13\pm 8.5$, and $g_{\rm NL}/10^5=10\pm 1.76$ against the
input of $10\pm 1.5$. We find that our method well reproduces the
input values of NG parameters and the error estimations using the
Fisher matrix (eq.[\ref{eq:fisher}]) provide reasonable measurement
error. The mean value of $\tau_{\rm NL}$ is found to be underestimated
by 5\%, which may be due to the incompleteness of the theoretical
estimation of the kurtosis parameters. The systematic error is, however, much
smaller than the statistical error of $\tau_{\rm NL}$ and the effect
on the final result is small.

\subsection{Comparison of constraints from MFs and skewness/kurtosis}
\begin{figure}
\caption{Distribution of the bestfit values of $f_{\rm NL}^{\rm
    (loc)}$ around the input value $f_{\rm NL}^{\rm input}=100$ using
  1000 NG simulation maps. We obtain the bestfit values using two
  different measurements: MFs (red lines) and three skewness
  parameters (blue lines). The differences between these two
  measurements are also plotted with the yellow lines.}
\begin{center}
\includegraphics[width=8cm]{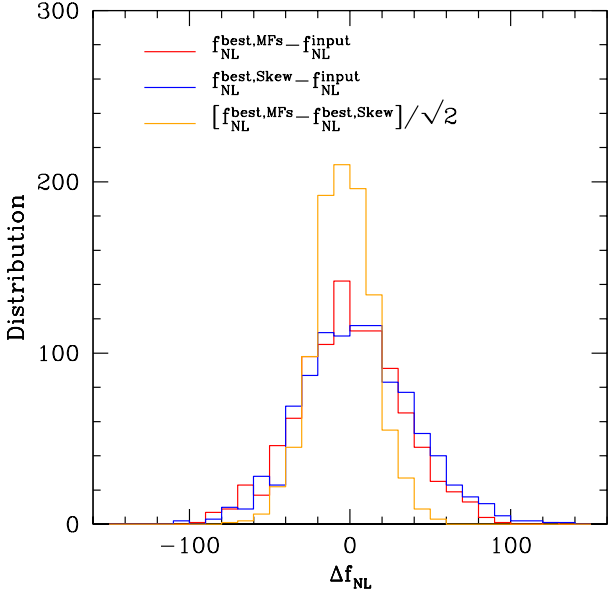}
\label{fig:mom_vs_mfs}
\end{center}
\end{figure}

Perturbative corrections of MFs are determined by skewness
(eq:[\ref{eq:skewness}]) and kurtosis parameters
(eq.[\ref{eq:kurtosis}]) upto 2nd order. The parameters carry the NG
information and hence their measured values can be directly used to
limit the NG parameters.  For consistency checks, we compare the
constraints on NG parameters estimated from MFs with those estimated
directly from skewness and kurtosis parameters (hereafter we call them
``moments'') using WMAP mock simulation maps. Fig.
\ref{fig:mom_vs_mfs} shows the distribution of $f_{\rm NL}^{\rm
  (loc)}$ estimated from MFs (red), and those from skewness values
(blue). We find that the sample variance are comparable and thus MFs
and the moments have similar power to constrain $f_{\rm NL}^{(loc)}$. We
also plot the difference of the bestfit values of $f_{\rm NL}^{(loc)}$
from the MFs and the skewness divided $\sqrt{2}$. The limits from the MFs and
the moments do not completely agree because the weights on skewness
parameters in MFs are not equal. The difference is smaller than the
dispersion of the distribution of $f_{\rm NL}$, which means that the
measurements of MFs and moments are strongly correlated.

\section{Application to WMAP 7-year data}
\label{sec:observations}
\begin{figure*}
\caption{Three MFs for WMAP 7-year temperature maps at different
  $\theta$=40, 20, 10, 7, and 5 arcmin from the top to bottom. The Gaussian
  term $V_k^{\rm (G)}$ is subtracted to focus on the deviation from
  Gaussian. For reference, the first and second-order perturbative
  corrections with the bestfit values of $f_{\rm NL}^{\rm (loc)}$,
  $\tau_{\rm NL}$ and $g_{\rm NL}$ are plotted, respectively.}
\begin{center}
\includegraphics[width=16cm]{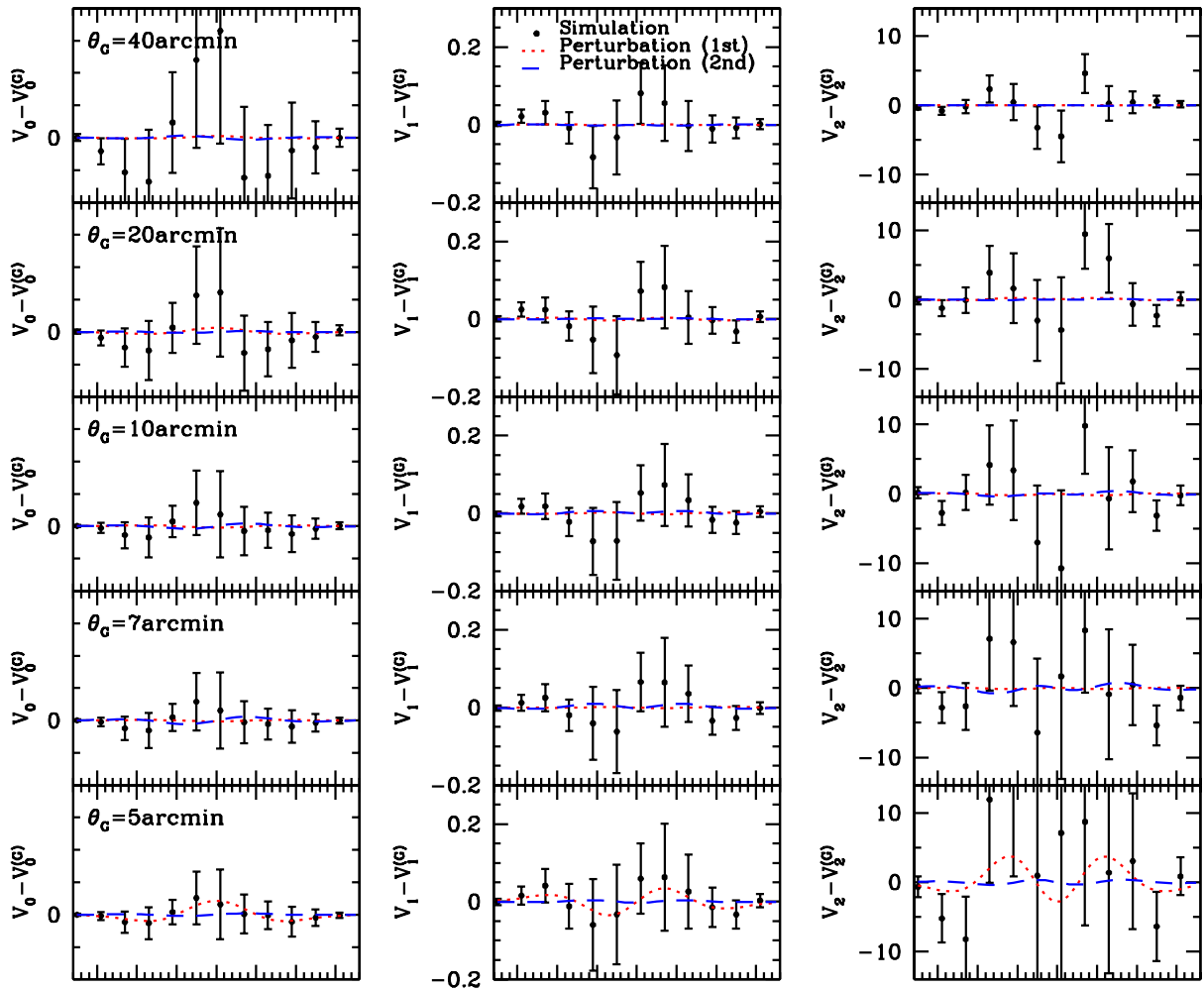}
\label{fig:mfobs}
\end{center}
\end{figure*}

\begin{figure}
\caption{Three skewness and four kurtosis parameters measured from WMAP
  7-year data at different smoothing scales of $\theta$. The
  definitions of the skewness and kurtosis parameters are given in the equations
  (\ref{eq:skewness}) and (\ref{eq:kurtosis}), respectively.}
\begin{center}
\includegraphics[width=8cm]{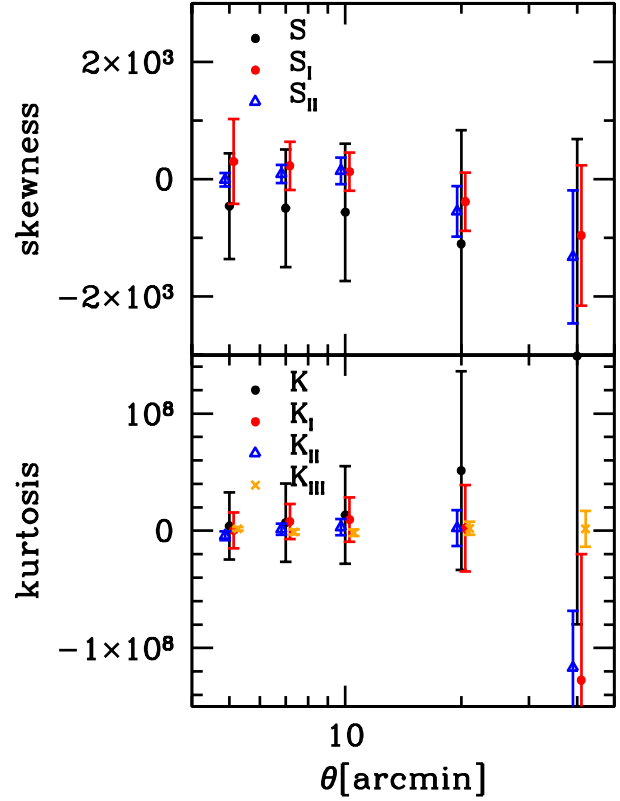}
\label{fig:momobs}
\end{center}
\end{figure}

In the previous section, we show that the perturbation works even
including the various observational effects.  We apply the
perturbative formulae to WMAP 7-year temperature maps and give limits
on the NG parameters. Fig. \ref{fig:mfobs} shows the MFs of WMAP
7-year V+W co-added maps at different smoothing scales.  We subtract
Gaussian term $V_k^{\rm (G)}$ from the observed MFs to focus on the deviation
from Gaussian. For comparison, we plot the perturbative formulae with
the bestfit values of $f_{\rm NL}^{\rm (loc)}, \tau_{\rm NL}$, and
$g_{\rm NL}$. The 1st-order ($f_{\rm NL}^{\rm (loc)}$) and 2nd-order
($\tau_{\rm NL}$ and $g_{\rm NL}$) contributions are shown with dotted
and dashed line, respectively. The differences are consistent with zero
for all MFs at different smoothing scales. Fig. \ref{fig:momobs}
shows the skewness and kurtosis parameters of the same WMAP data at
different smoothing scales, which are also consistent with zero.

Table \ref{tab:limit_fnl} lists the limits on $f_{\rm NL}$s,
$\tau_{\rm NL}$ and $g_{\rm NL}$.  The constraining power is strongest
at $\theta$=10 or 7 arcmin scales because the noise is dominated at
smaller scales. Maps of different smoothing scales have different
scale information and hence combining results from different $\theta$
maps provide stronger constraint.  These constraints take into account
the point source effect and its contribution is marginalized over. All
types of primordial NG parameters we consider are consistent with
zero. The constraints from MFs are weaker than the optimal estimator
based on the bispectrum \citep{Komatsuetal:11} and on the trispectrum
\citep{Smidtetal:10}. This is because that the skewness and kurtosis
parameters lose configuration information on the bispectrum and the
trispectrum. Again we stress that the consistency check from MFs are
important to check the systematic effects. It may be interesting that
the bestfit value of $\tau_{\rm NL}$ from MFs is slightly inclined to
be negative because the negative value of $\tau_{\rm NL}$ is not
allowed for all of multi-field inflation models by the inequality
relation $\tau_{\rm NL}>36/25f_{\rm NL}^{\rm (loc) 2}$
\citep{SuyamaYamaguchi:08}.  This inclination is stronger before
marginalization of point source effect as seen in Table
\ref{tab:limit_fnl_before}, however the significancy is still very
small.  We also give the limits from different frequency bands listed
in Table \ref{tab:limit_band}. The differences between different
frequency maps are within 1 sigma of statistical error, which means
that the frequency-dependent systematics such as Galactic foreground
do not affect our results so much. The results from MFs and moments
(skewness and kurtosis) are also consistent and their difference is
within 1$\sigma$ statistical error.

\begin{table*}
\caption{Limits on NG parameters from WMAP 7-year V+W co-added maps at
  different smoothing scales $\theta$ and when combined. We consider
  five NG parameters: quadratic NGs in local type $f_{\rm NL}^{\rm
    (loc)}$, equilateral type $f_{\rm NL}^{\rm (eq)}$, orthogonal type
  $f_{\rm NL}^{\rm (ort)}$, and cubic NGs of $g_{\rm NL}$, and
  $\tau_{\rm NL}$. We list the constraints from MFs in the upper part
  and moments (i.e., three skewness and four kurtosis parameters) in
  the lower part of the table. The point source effect is marginalized
  over for all these limits.}
\begin{center}
\begin{tabular}{ccccccc}
\hline\hline
measurement & $\theta$ & $f_{\rm NL}^{\rm (loc)} $ & $f_{\rm NL}^{\rm (eq)}$ & $f_{\rm NL}^{\rm (ort)}$ & $\tau_{\rm NL}/10^4$ & $g_{\rm NL}/10^5$ \\
\hline
MFs & 100  & $  -207 \pm   296$ & $ -2860 \pm  5540$ & $ 260 \pm   523$ & $  -180\pm 196$ & $  10 \pm  28 $\\
    &  70  & $  -117 \pm   173$ & $ -7400 \pm  4360$ & $  -2 \pm   399$ & $  -86\pm 105$ & $   0.7 \pm  18 $\\
    &  40  & $    14 \pm    86$ & $ -1350 \pm  3930$ & $-119 \pm   261$ & $  -24\pm  41$ & $   3.5 \pm  11 $\\
    &  20  & $     6 \pm    54$ & $  -361 \pm  2400$ & $ -58 \pm   204$ & $  -0.3\pm  17$ & $   0.8 \pm   8.1 $\\
    &  10  & $     3 \pm    53$ & $   -21 \pm   585$ & $ -25 \pm   195$ & $  -2.9\pm  11$ & $  -1.1 \pm   7.6 $\\
    &  7   & $    39 \pm    83$ & $  -354 \pm   768$ & $-118 \pm   252$ & $  -6.2\pm  11$ & $  -3.9 \pm   8.0 $\\
    &  5   & $    82 \pm   127$ & $  -879 \pm  1300$ & $-168 \pm   299$ & $  -3.0\pm  15$ & $  -1.1 \pm   9.8 $\\
 &  Combined & $    20 \pm    42$ & $ -121 \pm   208$ & $-129 \pm   171$ & $  -7.6\pm   8.7$ & $  -1.9 \pm   6.4 $\\
\hline
Moments  & 100 & $ -104 \pm   387$ & $-8240 \pm  5380$ & $-518 \pm 758$ & $ -216\pm 205$ & $  15.7\pm 44.3$\\
         & 70  & $ -163 \pm   204$ & $-6610 \pm  4200$ & $  22 \pm 518$ & $  -83\pm  99$ & $   9.9\pm 23.4$\\
         & 40  & $  -55 \pm    90$ & $-3840 \pm  3810$ & $  73 \pm 288$ & $  -43\pm  37$ & $   2.0\pm 11.4$\\
         & 20  & $  -18 \pm    55$ & $  715 \pm  2440$ & $  24 \pm 212$ & $    3.6\pm  14$ & $   1.8\pm  7.8$\\
         & 10  & $   15 \pm    54$ & $ -139 \pm   588$ & $ -77 \pm 203$ & $   -3.4\pm   9.6$ & $  -2.1\pm  7.1$\\
         & 7   & $   51 \pm    84$ & $ -436 \pm   770$ & $-165 \pm 261$ & $   -4.0\pm   9.3$ & $  -2.6\pm  7.6$\\
         & 5   & $   84 \pm   133$ & $ -846 \pm  1340$ & $-186 \pm 314$ & $   -2.2\pm  13$ & $  -0.5\pm  9.4$\\
   &  Combined & $   31 \pm    40$ & $ -132 \pm   196$ & $-145 \pm 178$ & $   -6.3\pm   8.0$ & $  -4.1\pm  5.8$\\
\hline
\end{tabular}
\label{tab:limit_fnl}
\end{center}
\end{table*}

\begin{table*}
\caption{Same as the combined limits in Table \ref{tab:limit_fnl} but
  for the limits before marginalization of point source effect.}
\begin{center}
\begin{tabular}{ccccccc}
\hline\hline
measurement & $\theta$ & $f_{\rm NL}^{\rm (loc)} $ & $f_{\rm NL}^{\rm (eq)}$ & $f_{\rm NL}^{\rm (ort)}$ & $\tau_{\rm NL}/10^4$ & $g_{\rm NL}/10^5$ \\
\hline
MFs & Combined & $    17 \pm    41$ & $ -129 \pm   198$ & $-129 \pm   171$ & $  -10.5\pm   8.1$ & $  -1.4\pm 6.4$ \\
Moments &  Combined & $   36 \pm    40$ & $ -77 \pm   188$ & $-154 \pm 177$ & $   -4.3\pm   7.6$ & $  -4.4\pm 5.8$ \\
\hline
\end{tabular}
\label{tab:limit_fnl_before}
\end{center}
\end{table*}

\begin{table*}
\caption{Limits on the NG parameters in different frequency bands.
  All of the limits are obtained from MFs (Upper) or Moments (Lower)
  by combining different $\theta_s$ maps and the point source effect
  is marginalized over.}
\begin{center}
\begin{tabular}{cccccccc}
\hline\hline
Estimator & band & $f_{\rm NL}^{\rm (loc)} $ & $f_{\rm NL}^{\rm (eq)}$ & $f_{\rm NL}^{\rm (ort)}$ & $\tau_{\rm NL}/10^4$ & $g_{\rm NL}/10^5$ \\
\hline
MFs    & Q+V+W & $ 22 \pm 43$ & $-185 \pm 211$ & $-216 \pm 172$ & $-10.6 \pm 8.9$ & $-1.9 \pm 6.3$ \\
       & Q     & $ 21 \pm 45$ & $  -5 \pm 264$ & $ -94 \pm 178$ & $-7.5 \pm 10.8$ & $-4.4 \pm 6.9$ \\
       & V     & $ 33 \pm 43$ & $ -61 \pm 220$ & $-143 \pm 174$ & $ -6.5 \pm 9.4$ & $-2.7 \pm 6.7$ \\
       & W     & $ 12 \pm 44$ & $-102 \pm 219$ & $ -98 \pm 174$ & $-7.5 \pm 10.8$ & $-4.4 \pm 6.9$ \\
\hline
Moments & Q+V+W & $ 25 \pm 41$ & $ -69 \pm 199$ & $ -90 \pm 181$ & $-10.4 \pm 8.1$ & $-4.8 \pm 5.8$ \\
       & Q     & $ 11 \pm 44$ & $ -48 \pm 241$ & $ -29 \pm 188$ & $-12.5 \pm 9.8$ & $-5.6 \pm 6.3$ \\
       & V     & $ 35 \pm 41$ & $ -62 \pm 206$ & $-118 \pm 180$ & $ -3.1 \pm 8.7$ & $-2.8 \pm 6.1$ \\
       & W     & $ 36 \pm 41$ & $ -91 \pm 206$ & $-137 \pm 179$ & $ -9.3 \pm 8.8$ & $-5.6 \pm 6.1$ \\
\hline
\end{tabular}
\label{tab:limit_band}
\end{center}
\end{table*}

\section{Summary and Discussions}

\label{sec:summary}
We first apply the perturbative formulae of MFs including second-order
NG to WMAP 7-year data and give limits on $\tau_{\rm NL}$ and $g_{\rm
  NL}$ as well as local-type, equilateral-type and orthogonal-type
$f_{\rm NL}$. Consistency check using different estimators are
important to obtain more robust results because different estimators
are sensitive to different aspects and systematics. We find no
evidence of NG from any type of NG components and then obtain the
limits on each NG parameter: $f_{\rm NL}^{\rm (loc)}=20\pm 42$,
$f_{\rm NL}^{\rm (eq)}=-121\pm 208$, and $f_{\rm NL}^{\rm
  (ort)}=-129\pm 171$, $\tau_{\rm NL}/10^4=-7.6\pm 8.7$ and $g_{\rm
  NL}/10^5=-1.9\pm 6.4$. Our result is consistent with the previous
works using the estimators of bispectrum and trispectrum.
Constraining $\tau_{\rm NL}$ is important for testing multiple
inflation models which must satisfy the inequality condition
$\tau_{\rm NL}>36/25 f_{\rm NL}^2$. Our limit on $\tau_{\rm NL}$ is
consistent with zero, but the bestfit value of $\tau_{\rm NL}$ is
inclined to be negative even after subtracting point source
effects. Upcoming CMB experiments such as Planck will give
statistically better results.

Large-scale structure offers another test to study the presence of
primordial NG. Scale-dependent bias of halo clustering also provides a
powerful probe of primordial NG \citep{Dalaletal:08} and observational
constraints $-29<f_{\rm NL}^{\rm (loc)}<70$ (95\%C.L.)  are obtained
\citep{Slosaretal:08}. \citet{DesjacquesSeljak:10} give constraints on
$g_{\rm NL}$ as $-3.5\times 10^5<g_{\rm NL}<8.2\times 10^5$ (95\%
C.L.)  from the halo mass function and halo bias.  Combining galaxy
bispectra is also useful to constrain the primordial NG including the
higher-order NGs $\tau_{\rm NL}$ and $g_{\rm NL}$
\citep{JeongKomatsu:09,Nishimichietal:10}.

The secondary effects like a coupling between the integrated
Sachs-Wolfe effect and the gravitational lensing
\citep{GoldbergSpergel:99} may contaminate the measurement; however,
the estimation is small of the order of $f_{\rm NL}\sim 3$
\citep{Komatsuetal:11}.  In this paper we give limits on the three
types of $f_{\rm NL}$s and $\tau_{\rm NL}$ and $g_{\rm NL}$ locally
given. In general, there is still a wide range of NG such as
isocurvature NG \citep{Kawasakietal:08,Hikageetal:09}. The equilateral
type of $g_{\rm NL}$ has been also given by
\citet{MizunoKoyama:10}. The application to the other types of NG is a
future work.


\section{Acknowledgments}
We acknowledge James Fergusson for carefully reviewing the manuscript
and providing very useful comments. We also thank Eiichiro Komatsu for
helpful comments. This work is supported by Grant-in-Aid for
Scientific Research from the Ministry of Education, Science, Sports,
and Culture of Japan No.\,24740160 (C.~H.).

{}


\begin{thebibliography}{99}

\bibitem[\protect\citeauthoryear{Alishahiha et al.}{2004}]{Alishahihaetal:04}
Alishahiha,~M., Silverstein,~E., Tong,~D., 2004, Phys. Rev. D, 70, 123505

\bibitem[\protect\citeauthoryear{Acquaviva et al.}{2003}]{Acq03}
Acquaviva~V., Bartolo~N., Matarrese~S., Riotto~A., 2003, Nuclear Phys. B, 667, 119

\bibitem[\protect\citeauthoryear{Arkani-Hamed et al.}{2004}]{ArkaniHamedetal:04}
Arkani-Hamed,~N., Creminelli,~P., Mukohyama,~S., Zaldarriaga,~M., 2004,
J. Cosmol. Astropart. Phys., 4, 1

\bibitem[\protect\citeauthoryear{Babich, Creminelli \& Zaldarriaga}{2004}]{Babichetal:04}
Babich~D., Creminelli~P., Zaldarriaga~M. 2004, J. Cosmol. Astropart. Phys., 8, 9

\bibitem[\protect\citeauthoryear{Bartolo, Matarrese \& Riotto}{2006}]{Bartolo06}
Bartolo~N., Matarrese~S., Riotto~A., 2006, J. Cosmol. Astropart. Phys., 6, 24

\bibitem[\protect\citeauthoryear{Buchbinder, Khoury \& Ovrut}{2007}]{Buch07}
Buchbinder~E.~I., Khoury~J., Ovrut~B.~A., 2007, J. High Energy Phys., 11, 76

\bibitem[\protect\citeauthoryear{Chen, Easther \& Lim}{2007}]{ChenEasterLim:07}
Chen~X., Easther~R., Lim~E.~A., 2007, J. Cosmol. Astropart. Phys., 6, 23

\bibitem[\protect\citeauthoryear{Creminelli, \& Senatore}{2007}]{CS2007}
Creminelli,~P., Senatore,~L., 2007, J. Cosmol. Astropart. Phys., 11, 10

\bibitem[\protect\citeauthoryear{Creminelli, Senatore, \& Zaldarriaga}{2007}]{Creminelli:07} 
Creminelli~P., Senatore~L., Zaldarriaga~M., 2007, J. Cosmol. Astropart. Phys., 3, 19 

\bibitem[\protect\citeauthoryear{Curto et al.}{2011}]{Curtoetal:11} 
Curto~A., Mart${\rm \acute{i}}$nez-Gonz${\rm \acute{a}}$lez~E., Barreiro~R.~B., Hobson~M.~P., 2011, MNRAS, 417, 488

\bibitem[\protect\citeauthoryear{Dalal et al.}{2008}]{Dalaletal:08}
Dalal,~N., Dor$\acute{e}$,~O., Huterer,~D., Shirokov,~A., 2008, Phys. Rev. D, 77, 123514

\bibitem[\protect\citeauthoryear{Desjacques \& Seljak}{2010}]{DesjacquesSeljak:10}
Desjacques,~V., Seljak,~U., 2010, Phys. Rev. D, 81, 023006

\bibitem[\protect\citeauthoryear{Dvali, Gruzinov \& Zaldarriaga}{2004}]{DGZ2004}
Dvali,~G., Gruzinov,~A., Zaldarriaga,~M., 2004, Phys. Rev. D, 69, 083505

\bibitem[\protect\citeauthoryear{Elsner \& Wandelt}{2009}]{ElsnerWandelt:09}
Elsner,~F., Wandelt,~B.~D., 2009, ApJS, 184, 264

\bibitem[\protect\citeauthoryear{Falk et al.}{1993}]{Falk93}
Falk~T., Madden~R., Olive~K.~A., Srednicki~M., 1993, Phys. Lett. B318, 354

\bibitem[\protect\citeauthoryear{Fergusson, Regan, Shellard}{2010}]{Fergussonetal:10}
Fergusson~J.~R., Regan~D.~M., Shellard~E.~P.~S., 2010, arXiv:1012.6039

\bibitem[\protect\citeauthoryear{Gangui et al.}{1994}]{Gangui94}
Gangui~A., Lucchin~F., Matarrese~S., Mollerach~S., 1994, ApJ, 430, 447

\bibitem[\protect\citeauthoryear{Gold et al.}{2011}]{Goldetal:11}
Gold.~B. et al., 2011, ApJS, 192, 15

\bibitem[\protect\citeauthoryear{Goldberg \& Spergel}{1999}]{GoldbergSpergel:99}
Goldberg~D.~M., Spergel~D.N., 1999, Phys. Rev. D, 59, 103002

\bibitem[\protect\citeauthoryear{Gupta et al.}{2002}]{GuBeHea02} 
Gupta~S., Berera~A., Heavens~A.~F., Matarrese S., 2002, Phys. Rev. D, 66, 043510

\bibitem[\protect\citeauthoryear{Hikage, Komatsu \& Matsubara}{2006}]{Hikageetal:06}
Hikage,~C., Komatsu,~E., Matsubara,~T., 2006, ApJ, 653, 11

\bibitem[\protect\citeauthoryear{Hikage et al.}{2008}]{Hikageetal:08}
Hikage,~C., Matsubara,~T., Coles,~P., Liguori,~M., Hansen,~F.~K.,
Matarrese,~S., 2008, MNRAS, 389, 1439

\bibitem[\protect\citeauthoryear{Hikage et al.}{2009}]{Hikageetal:09}
Hikage~C., Koyama~K., Matsubara~T., Takahashi~T., Yamaguchi~M., 2009, 
MNRAS, 398, 2188

\bibitem[\protect\citeauthoryear{Jeong \& Komatsu}{2009}]{JeongKomatsu:09}
Jeong,~D., Komatsu,~E., 2009, ApJ, 703, 1230

\bibitem[\protect\citeauthoryear{Kawasaki et al.}{2008}]{Kawasakietal:08}
Kawasaki,~M., Nakayama,~K., Sekiguchi,~T., Suyama,~T., Takahashi,~F.,
2008, JCAP, 11, 19

\bibitem[\protect\citeauthoryear{Komatsu \& Spergel}{2001}]{KomatsuSpergel:01}
Komatsu,~E., Spergel,~D.~N., 2001, Phys. Rev. D, 63, 63002

\bibitem[\protect\citeauthoryear{Komatsu et al.}{2011}]{Komatsuetal:11}
Komatsu,~E. et al., 2011, ApJS, 192, 18

\bibitem[\protect\citeauthoryear{Koyama et al.}{2007}]{Koya07}
Koyama~K., Mizuno~S., Vernizzi~F., Wands~D., 2007, J. Cosmol. Astropart. Phys., 11, 24

\bibitem[\protect\citeauthoryear{Kogo \& Komatsu}{2006}]{KogoKomatsu:06}
Kogo,~N., Komatsu,~E., 2006, Phys. Rev. D, 73, 083007

\bibitem[\protect\citeauthoryear{Linde \& Mukhanov}{1997}]{LM1997}
Linde~A.~D., Mukhanov~V., 1997, Phys. Rev. D, 56, R535

\bibitem[\protect\citeauthoryear{Lim \& Simon}{2012}]{LimSimon:12}
Lim,~E.~A., Simon,~D., 2012, J. Cosmol. Astropart. Phys., 1, 48

\bibitem[\protect\citeauthoryear{Lyth, Ungarelli \& Wands}{2003}]{Lyth03}
Lyth~D.~H., Ungarelli~C., Wands~D., 2003, Phys. Rev. D, 67, 023503

\bibitem[\protect\citeauthoryear{Maldacena}{2003}]{Mal03}
Maldacena~J.~M., 2003, J. High Energy Phys., 5, 13

\bibitem[\protect\citeauthoryear{Matsubara}{2003}]{Matsubara:03}
Matsubara,~T., 2003, ApJ, 584, 1

\bibitem[\protect\citeauthoryear{Matsubara}{2010}]{Matsubara:10}
Matsubara,~T., 2010, Phys. Rev. D, 81, 083505

\bibitem[\protect\citeauthoryear{Mecke et al.}{1994}]{MeckeBuchertWagner:94}
Mecke,~K.~R., Buchert,~T., \& Wagner,~H. 1994, A\&A, 288, 697

\bibitem[\protect\citeauthoryear{Mizuno \& Koyama}{2010}]{MizunoKoyama:10}
Mizuno,~S., Koyama,~K., 2010, J. Cosmol. Astroprt. Phys., 10, 2

\bibitem[\protect\citeauthoryear{Moss \& Xiong}{2007}]{MossXiong:07}
Moss~I., Xiong~C., 2007, J. Cosmol. Astropart. Phys., 4, 7

\bibitem[\protect\citeauthoryear{Natoli et al.}{2010}]{Natolietal:10}
Natoli~P., et al., 2010, MNRAS, 408, 1658

\bibitem[\protect\citeauthoryear{Nishimichi et al.}{2010}]{Nishimichietal:10}
Nishimichi,~T., Taruya,~A., Koyama,~K., Sabiu,~C., J. Cosmol. Astropart. Phys., 7, 2

\bibitem[\protect\citeauthoryear{Okamoto \& Hu}{2002}]{OkamotoHu:02}
Okamoto,~T., Hu,~W., 2002, Phys. Rev. D, 66, 63008

\bibitem[\protect\citeauthoryear{Salopek \& Bond}{1990}]{Salopek90} 
Salopek~D.~S., Bond~J.~R., 1990, Phys. Rev. D, 42, 3936 

\bibitem[\protect\citeauthoryear{Seery \& Lidsey}{2005}]{SeeryLidsey:05} 
Seery~D., Lidsey~J.~D., 2005, J. Cosmol. Astropart. Phys., 6, 3

\bibitem[\protect\citeauthoryear{Schmalzing \& Buchert}{1997}]{SchmalzingBuchert:97} 
Schmalzing,~J., \& Buchert,~T., 1997, ApJ, 482, L1

\bibitem[\protect\citeauthoryear{Senatore, Tassev, Zaldarriaga}{2009}]{SenatoreTassevZaldarriaga:09} 
Senatore,~L., Tassev,~S., Zaldarriaga,~M. 2009, J. Cosmol. Astropart. Phys., 8, 31

\bibitem[\protect\citeauthoryear{Slosar et al.}{2008}]{Slosaretal:08}
Slosar,~A., Hirata,~C., Seljak,~U., Ho,~S., Padmanabhan,~N., 
2008, J. Cosmol. Astropart. Phys., 8, 31

\bibitem[\protect\citeauthoryear{Smidt et al.}{2010}]{Smidtetal:10}
Smidt,~J., Amblard,~A., Byrnes,~C.~T., Cooray,~A., Heavens,~A., Munshi,~D., 
2010, Phys. Rev. D, 81, 123007

\bibitem[\protect\citeauthoryear{Suyama et al.}{2010}]{STYY:10}
Suyama,~T., Takahashi,~T., Yamaguchi,~M., Yokoyama,~S., 
2010, J. Cosmol. Astropart. Phys., 12, 30

\bibitem[\protect\citeauthoryear{Suyama \& Yamaguchi}{2008}]{SuyamaYamaguchi:08}
Suyama,~T., Yamaguchi,~M., 2008, Phys. Rev. D, 77, 023505

\bibitem[\protect\citeauthoryear{Vielva \& Sanz}{2010}]{VielvaSanz:10}
Vielva,~P., Sanz,~J.~L., 2010, MNRAS, 404, 895

\bibitem[\protect\citeauthoryear{Yadav \& Wandelt}{2008}]{YadavWandelt:08}
Yadav,~A.~P.~S., Wandelt,~B.,~D., 2008, Phys. Rev. Lett., 100, 181301


\end{thebibliography}
\end{document}